\documentclass[preprint]{aastex}

\usepackage{graphicx}
\usepackage{txfonts}
\usepackage{natbib}
\usepackage{longtable}

\newcommand{\hts}{H$_{2}$S}
\newcommand{\htsi}{H$_{2}$$^{32}$S}
\newcommand{\htsii}{H$_{2}$$^{33}$S}
\newcommand{\htsiii}{H$_{2}$$^{34}$S}
\newcommand{\hto}{H$_{2}$O}

\newcommand{\cmc}{cm$^{-3}$}
\newcommand{\cms}{cm$^{-2}$}

\newcommand{\ntot}{$N_{\rm tot}$}
\newcommand{\nobs}{$N_{\rm obs}$}
\newcommand{\nht}{$n_{\rm H2}$}
\newcommand{\tkin}{$T_{\rm kin}$}
\newcommand{\eup}{$E_{\rm up}$}
\newcommand{\trot}{$T_{\rm rot}$}
\newcommand{\nup}{$N_{\rm u}$}
\newcommand{\vlsr}{v$_{\rm lsr}$}
\newcommand{\dv}{$\Delta$v}
\newcommand{\tex}{$T_{\rm ex}$}
\newcommand{\um}{$\mu$m}
\newcommand{\tmb}{$T_{\rm mb}$}
\newcommand{\tpeak}{$T_{\rm peak}$}
\newcommand{\tint}{$\int{T_{\rm mb}d\rm v}$}
\newcommand{\chisq}{$\chi^{2}$}
\newcommand{\redchisq}{$\chi_{\rm red}^{2}$}
\newcommand{\amm}{NH$_{3}$}

\newcommand{\hh}{H$_{2}$}
\newcommand{\sn}{~$\times$~10}
\newcommand{\nhh}{$N_{\rm H_{2}}$}
\newcommand{\mc}{CH$_{3}$CN}
\newcommand{\hctn}{HC$_{3}$N}
\newcommand{\tdust}{$T_{\rm dust}$}
\newcommand{\taudust}{$\tau_{\rm d}$}
\newcommand{\taudustnu}{$\tau_{\rm d}(\nu)$}
\newcommand{\ssize}{$\theta_{\rm s}$}
\newcommand{\tauiso}{$\tau_{\rm iso}$}

\newcommand{\herschel}{\emph{Herschel}}
\newcommand{\mhyd}{$\rm m_{H}$}
\newcommand{\chidust}{$\chi_{\rm dust}$}
\newcommand{\opaco}{$\kappa_{\rm 1.3mm}$}
\newcommand{\fbp}{C13}
\newcommand{\bsize}{$\theta_{\rm b}$}
\newcommand{\beamff}{$\eta_{\rm bf}$}
\newcommand{\nsi}{$N_{i}$}

\shorttitle{\hts\ toward the Orion KL hot core}
\shortauthors{Crockett et al.}

\begin{document}

\title{\emph{Herschel} observations of EXtra-Ordinary Sources: \hts\ AS A PROBE OF DENSE GAS AND POSSIBLY HIDDEN LUMINOSITY TOWARD THE ORION KL HOT CORE\footnote{Herschel is an ESA space observatory with science instruments provided by European-led Principal Investigator consortia and with important participation from NASA.}}

\author{N.~R.~Crockett, E.~A.~Bergin, J.~L.~Neill}
\affil{Department of Astronomy, University of Michigan, 500 Church Street, Ann Arbor, MI 48109, USA}
\author{J.~H.~Black}
\affil{Department of Earth and Space Sciences, Chalmers University of Technology, Onsala, Sweden}
\author{G.~A.~Blake, and M.~Kleshcheva}
\affil{California Institute of Technology, Division of Geological and Planetary Sciences, MS 150-21, Pasadena, CA 91125, USA}

\begin{abstract}
We present \emph{Herschel}/HIFI observations of the light hydride \hts\ obtained from the full spectral scan of the Orion Kleinmann-Low nebula (Orion KL) taken as part of the HEXOS GT key program. In total, we observe 52, 24, and 8 unblended or slightly blended features from \htsi, \htsiii, and \htsii, respectively. We only analyze emission from the so called hot core, but emission from the plateau, extended ridge, and/or compact ridge are also detected. Rotation diagrams for ortho and para \hts\ follow straight lines given the uncertainties and yield \trot~=~141~$\pm$~12~K. This indicates \hts\ is in LTE and is well characterized by a single kinetic temperature or an intense far-IR radiation field is redistributing the population to produce the observed trend. We argue the latter scenario is more probable and find that the most highly excited states (\eup~$\gtrsim$~1000~K) are likely populated primarily by radiation pumping. We derive a column density, \ntot(\htsi)~=~9.5~$\pm$~1.9~$\times$~10$^{17}$~\cms, gas kinetic temperature, \tkin~=~120$\pm^{13}_{10}$~K, and constrain the \hh\ volume density, \nht~$\gtrsim$~9\sn$^{7}$~\cmc, for the \hts\ emitting gas. These results point to an \hts\ origin in markedly dense, heavily embedded gas, possibly in close proximity to a hidden self-luminous source (or sources), which are conceivably responsible for Orion KL's high luminosity. We also derive an \hts\ ortho/para ratio of 1.7~$\pm$~0.8 and set an upper limit for HDS/\hts\ of $<$~4.9\sn$^{-3}$.
\end{abstract}

\section{Introduction}
\label{s-intro}

The sub-mm and far-IR are fruitful parts of the electromagnetic spectrum to study light hydride molecules. Due to their low molecular weight, the rotation transitions of these species are more widely spaced and occur at higher frequencies than their more complex counterparts. Sub-mm and mm wave observations show that light hydrides (e.g. H$_{2}$O, NH$_{3}$, HCl, \hts, etc...) are common in the interstellar medium \citep[ISM;][]{phillips87}. However, the use of these molecules as physical probes has been hindered primarily by atmospheric absorption. Although many light hydrides have low lying rotation (or inversion) transitions at wavelengths that are accessible through open atmospheric windows, observations of higher energy lines occur at frequencies $>$~1~THz, where the atmosphere is completely opaque. As a result, unambiguous constraints on molecular emissions of many key light hydrides are rare. The HIFI instrument \citep{degraauw10} on board the \emph{Herschel} Space Observatory \citep{pilbratt10}, however, provides the first opportunity to access this part of the electromagnetic spectrum at high spectral resolution, making light hydrides available as physical probes of molecular gas. 

In this study, we investigate and model the emission of \hts\ toward the hot core within the Orion Kleinmann-Low nebula (Orion KL), the paradigmatic massive star forming region in our Galaxy. Historically \hts\ has been used primarily as a probe of sulfur chemistry in the ISM. A number of studies have measured \hts\ abundances toward a wide variety of environments including dark clouds \citep{minh89}, low density molecular clouds \citep{tieftrunk94}, low mass protostars \citep{buckle03}, hot cores \citep{hatchell98, vandertak03, herpin09}, and shocks \citep{minh90, minh91}. In this study, however, we explore the utility of \hts\ more generally as a probe of the gas physical state. As a light hydride, \hts\ has a high dipole moment (0.97~D) and widely spaced energy states. Furthermore, many transitions have critical densities in excess of 10$^{7-8}$~\cmc. Consequently, \hts\ is very sensitive to both the gas temperature and density. 

The data presented here were taken from the full \emph{Herschel}/HIFI spectral scan of Orion~KL obtained as part of the guaranteed time key program entitled \emph{Herschel Observations of EXtra Ordinary Sources} (HEXOS). Because of the unprecedented frequency coverage provided by this dataset ($\sim$1.2~THz), we had access to over 90 transitions from \htsi\ and its two rarer isotopologues \htsii\ and \htsiii\ over an energy range of 55~--~1233~K, many of which can not be observed from the ground because they occur at frequencies higher than 1~THz. With this comprehensive dataset, we are able to explore the viability of this molecule as a probe of the gas physical state and set direct constraints on the abundance of \hts\ toward the Orion hot core. 

Although we seek only to model the \hts\ hot core emission because it dominates the line profiles at high excitation energies, Orion~KL harbors several other spatial/velocity components \citep{blake87}. Despite the fact that these components are not spatially resolved by \emph{Herschel}, they have substantially different central velocities relative to the Local Standard of Rest, \vlsr, and full width at half maximum line widths, \dv. We can therefore differentiate these components using the spectral resolution of HIFI. These components include the already mentioned ``hot core" (\vlsr~$\approx$~3--5~km/s, \dv~$\approx$~5--10~km/s); at least two outflow components often referred to as the ``plateau" (\vlsr~$\approx$~7 -- 11~km/s, \dv~$\gtrsim$20~km/s); a group of dense clumps adjacent to the hot core collectively known as the ``compact ridge" (\vlsr~$\approx$~7--9~km/s, \dv~$\approx$~3--6~km/s); and widespread cool, quiescent gas referred to as the ``extended ridge" (\vlsr~$\sim$~9~km/s, \dv~$\sim$~4~km/s).

In specifying these components, we note that recent studies have questioned the designation of the so called Orion ``hot core" as a \emph{bona fide} hot core. \citet{zapata11}, for example, conclude that the Orion hot core is actually externally heated by an ``explosive event", possibly a stellar merger. In this scenario, the hot core was a relatively dense region within the extended ridge that has been further compressed by the flow of material produced by this event. \citet{goddi11}, on the other hand, find that the heavily embedded object radio source I may be the primary heating source for the Orion hot core. They suggest that the combined effect of source I's proper motion and outflow could be mechanically heating the gas and dust. In order to be consistent with the literature, we refer to this region as the hot core, but recognize that this may, indeed, be a misnomer. 

The structure of this paper is as follows. We detail our observations and data reduction procedure in Sec.~\ref{s-obs}. In Sec.~\ref{s-res} we present the results from our rotation diagram and non-LTE analyses. We also compute an ortho/para ratio and a D/H ratio upper limit for \hts. We discuss our results in Sec.~\ref{s-disc}, and summarize our conclusions in Sec.~\ref{s-con}.

\section{Observations and Data Reduction}
\label{s-obs}
The \hts\ transitions are scattered throughout the full HIFI spectral scan of Orion KL. More details on the HEXOS key program as well as HIFI spectral scans in general are given in \citet{bergin10}. A more comprehensive description of the data product is presented in \citet[][hereafter \fbp]{crockett13}. As part of a global analysis, \fbp\ also presents a spectral model for each detected molecule in the Orion~KL HIFI survey assuming local thermodynamic equilibrium (LTE). The sum of all of these fits yields the total modeled molecular emission, which we refer to, in this study, as the ``full band model".  We, however, briefly describe the entire dataset here and outline the data reduction process. Most of the observations were obtained in March and April 2010, with the exceptions of bands 3a and 5b, which were obtained 9 September 2010 and 19 March 2011, respectively. The full dataset covers a significant bandwidth of approximately 1.2 THz in the frequency range 480--1900~GHz, with gaps between 1280--1430~GHz and 1540--1570~GHz. The data have a spectral resolution of 1.1~MHz corresponding to 0.2~--~0.7~km/s across the band. The spectral scans for each band were taken in dual beam switch (DBS) mode, the reference beams lying 3\arcmin\ east or west of the target, using the wide band spectrometer (WBS) with a redundancy of 6 or 4 for bands 1--5 and 6--7, respectively -- see \citet{bergin10} for a definition of redundancy. The beam size, \bsize, of \herschel\ varies between 11\arcsec\ and 44\arcsec\ over the HIFI scan. For bands 1--5, the telescope was pointed toward coordinates $\alpha_{J2000} = 5^h35^m14.3^s$ and $\delta_{J2000} = -5^{\circ}22'33.7''$, midway between the Orion hot core and compact ridge. For bands 6--7, where the beam size is smaller, the telescope was pointed directly toward the hot core at coordinates $\alpha_{J2000} = 5^h35^m14.5^s$ and $\delta_{J2000} = -5^{\circ}22'30.9''$. We assume the nominal absolute pointing error (APE) for \emph{Herschel} of 2.0\arcsec\ \citep{pilbratt10}. 

The data were reduced using the standard HIPE \citep{ott10} pipeline version 5.0 (CIB 1648) for both the horizontal, H, and vertical, V, polarizations. Level 2 double sideband (DSB) scans, the final product of pipeline reduction, required additional processing before they could be deconvolved to a single sideband (SSB) spectrum. First, each DSB scan was inspected for spurious spectral features (``spurs") not identified by the pipeline. Once identified, these features were flagged by hand such that they would be ignored by the deconvolution routine. Baselines were also removed from the DSB scans using the FitBaseline task within HIPE by fitting polynomials to line free regions. In most instances we required a constant or first order polynomial; however, second order polynomials were used in some cases. 

Once the baselines were removed, we deconvolved the DSB scans to produce an SSB spectrum for each band. The deconvolution was performed using the \emph{doDeconvolution} task within HIPE. We did not apply a gain correction or use channel weighting. The deconvolution was done in three stages. First, the data were deconvolved with the strongest lines (T$_{A}^{*}$~$>$~10~K) removed. This reduced the probability of strong ghost lines appearing in the SSB spectrum \citep[see e.g.][]{comito02} but resulted in channel values of zero at strong line frequencies in the SSB spectrum. Second, we performed another deconvolution with the strongest lines present to recover the data at strong line frequencies. And third, the strong lines were incorporated into the the weak line deconvolution by replacing the zero value channels with the strong line deconvolution. 

The SSB spectra were exported from HIPE to FITS format using \emph{HiClass}. These files were then imported to CLASS{\footnote{http://www.iram.fr/IRAMFR/GILDAS}  and converted to CLASS format. All subsequent data reduction procedures were performed using this program. The H and V polarizations were averaged together and aperture efficiency corrections were applied using Eq.~1 and 2 from \citet{roelfsema12} with the reference wavelength equal to the central wavelength of each band. For bands~1--5, we use the aperture efficiency correction, which is more coupled to a point source, because the \emph{Herschel} beam size (\bsize~$\approx$~17\arcsec\ -- 44\arcsec) is large relative to the size of the hot core ($\sim$~3\arcsec\ -- 10\arcsec). For bands~6 -- 7 (\bsize~$\approx$~11\arcsec\ -- 15\arcsec), on the other hand, we use the main beam efficiency, because it is more coupled to an extended source. For simplicity, however, we will refer to line intensities as main beam temperatures, \tmb, regardless of the band in which a line falls. The H/V averaged, efficiency corrected, SSB spectra represent the final product of our data reduction procedure. 

\section{Results}
\label{s-res}
\subsection{Measuring Line Intensities}
\label{s-mli}

We identify 70 transitions of \htsi, the main isotopologue of \hts, spanning a range in upper state energy, \eup, of 55--1233~K. We also detect emission from the two rarer isotopologues \htsiii\ and \htsii, the former being more abundant than the latter. From these rarer species, we identify 44 and 21 transitions of \htsiii\ and \htsii\ spanning ranges in \eup\ of 55--700~K and 55--328~K, respectively. Fig.~\ref{p-ls} plots a sample of nine \htsi\ transitions over the full energy range over which emission is detected. Each panel plots a different transition, labeled in the upper left hand corner, with \eup\ (also labeled) increasing from upper left to lower right. From the figure, we see that the profiles of the most highly excited lines (\eup~$\gtrsim$~700~K) are simple and consistent with emission from the hot core. At lower excitation, however, the line profiles become more complex indicating emission from additional components. The wide component (\dv~$\sim$~30~km/s) clearly originates from the plateau, while the narrow component (\dv~$\sim$~3~km/s) is consistent with either the extended or compact ridge. 

Fig.~\ref{p-ls} also overlays Gaussian fits to each component, from which we obtain the \vlsr, \dv, peak line intensity, \tpeak, and integrated intensity, \tint, for each spatial/velocity component. We used CLASS to fit the observed line profiles after uniformly smoothing the data to $\sim$0.5~km/s, in order to increase the S/N, and fitting a local baseline. This was a relatively straight forward process at high energies because the hot core was the only emissive component. At lower energies, however, where multiple components were visible, different combinations of Gaussians could reproduce the observed profiles. In most instances, we allowed all of the parameters to vary during the fitting process. Although, care was taken to make sure that the Gaussian fits conformed to canonical values for \vlsr\ and \dv\ typical of the hot core, plateau, and extended ridge.  In rare cases, allowing all of the parameters to vary resulted in fits not in line with these known spatial/velocity components and the \vlsr\ was held fixed to ensure a reasonable fit.

Because Orion KL is an extremely line rich source in the sub-mm, a fraction of the potentially detected lines had to be excluded from our analysis due to strong line blends from other molecules. Blended \hts\ lines were identified by overlaying the full band model, which includes emission from all other identified species in the Orion KL HIFI scan (\fbp), to those spectral regions where we observe \hts\ emission. We split the observed lines into three categories: (1) lines that were not blended or any blending line predicted by the full band model had a peak flux $\lesssim$~10\% that of \hts, (2) lines that were blended but could be separated by Gaussian fitting, and (3) heavily blended lines from which reliable Gaussian fits could not be derived. We emphasized that these categories were determined by eye. We only include transitions from categories 1 and 2 in our analysis which amount to 52, 24, and 8 transitions from \htsi, \htsiii, and \htsii, respectively. We report the results of our Gaussian fits for the hot core, plateau, and extended/compact ridge in Tables~\ref{t-hc}, \ref{t-pl}, and \ref{t-er}, respectively for categories 1 and 2 defined above. As stated in Sec.~\ref{s-intro}, we seek only to investigate the hot core emission in this work, but we include our measurements for the extended and/or compact ridge and plateau here for completeness. Instances when the \vlsr\ was held fixed for a particular component during Gaussian fitting are indicated in Tables~\ref{t-hc} -- \ref{t-er}. We note that several Gaussian fits to the plateau have \vlsr\ values $<$~6~km/s, lower than what we would expect from this spatial/velocity component. These unusual line centers, marked in Table~\ref{t-pl}, are the result of nearby line blends, irregular baselines, weak emission from the plateau, or possible unidentified line blends. As a result, plateau line parameters for these transitions should be viewed with caution.

Fig.~\ref{p-gd} is an energy level diagram for \htsi. The lines connecting the levels are black, blue, or red corresponding, respectively, to categories 1, 2, and 3. The red transitions, therefore, are detected but do not provide any useful information for the present study because these lines are so blended that a reliable Gaussian fit could not be derived. Examination of this diagram reveals two things. First, that in spite of the prevalence of blending, we still are able to extract useful flux information over the energy range that \hts\ emission is detected. And second, that our primary probe of \hts\ for \eup~$\gtrsim$~350~K are the $\Delta$J~=~0, $\Delta$K$_{\pm}$~=~$\pm$1 transitions, and for \eup~$\lesssim$~350~K we become sensitive to $\Delta$J~=~$\pm$1 lines. Here, J is the quantum number corresponding to the total rotational angular momentum of \hts, while K$_{+}$ and K$_{-}$ are the quantum numbers that would correspond to the angular momentum along the axis of symmetry in the limit of an oblate or prolate symmetric top, respectively. Unfortunately, we do not detect many of the J~=~9 para transitions because they either occur in HIFI frequency coverage gaps or regions of high noise. 

\subsection{Measuring Upper State Column Densities}
\label{s-nup}
 
As a consequence of HIFI's wide frequency coverage, we directly measure \hts\ upper state column densities, \nup, for a large fraction of the available states. For more detailed derivations of the equations that follow and discussions of their utility in the analysis of molecular spectroscopic data see e.g. \citet{goldsmith99} and \citet{persson07}.  The analysis below requires the use of known isotopic ratios. We have carried out these calculations assuming two different sets of ratios. The first assumes solar values, $^{32}$S/$^{34}$S~=~22 and $^{32}$S/$^{33}$S~=~125 \citep{asplund09}, and the second were measured directly toward Orion~KL, $^{32}$S/$^{34}$S~=~20 and $^{32}$S/$^{33}$S~=~75 \citep{tercero10}. The text that follows refers to these sets of values as ``Solar" and ``Orion~KL" isotopic ratios, respectively. 

In instances where the main isotopologue, \htsi, is optically thick but we observe the same transition with \htsiii\ or \htsii, we compute the upper state column by explicitly calculating an optical depth. The optical depth can be estimated assuming that the excitation temperature is the same for both isotopologues and that $\tau_{main}$~$\gg$~$\tau_{iso}$, where $\tau_{main}$ is the optical depth of the \htsi\ line and $\tau_{iso}$ is the optical depth of the \htsii\ or \htsiii\ line. With these assumptions we can write the following relation,
\begin{equation}
\tau_{iso}=-ln\left(1 - \frac{T_{iso}}{T_{main}}  \right)
\end{equation}
where T$_{main}$ and T$_{iso}$ are the peak line intensities for the main and rarer isotopologue, respectively. The \htsi\ optical depth can then be computed using,
\begin{equation}
\tau_{main} = \tau_{iso} \left( \frac{^{32}S}{^{iso}S} \right).
\end{equation} 
The upper state column can then be computed using the relation,
\begin{equation}
N_{u}\,(H_{2}^{32}S)= \frac{8\pi k \nu^{2}}{hc^{3}A_{ul}}\frac{\int{T_{mb}d \rm v}}{\eta_{\rm bf}} \left(\frac{\tau_{main}}{1-e^{-\tau_{main}}} \right)  e^{\tau_{d}(\nu)}
\label{e-nup_thick}
\end{equation}
where $\nu$ is the rest frequency of the transition, \tint\ is the integrated intensity of the line in velocity space, \beamff\ is the beam filling factor, $^{32}$S/$^{iso}$S is the appropriate isotopic ratio, \taudustnu\ is the dust optical depth, A$_{ul}$ is the Einstein coefficient for spontaneous emission, $h$ is Planck's constant, $k$ is Boltzmann's constant, and $c$ is the speed of light. In Eq.~\ref{e-nup_thick}, the beam filling factor is given by the usual expression assuming a gaussian profile for both the telescope beam and source,
\begin{equation}
\eta_{bf}=\frac{\theta_{s}^{2}}{\theta_{s}^{2}+\theta_{b}^{2}}
\label{e-bff}
\end{equation}
where $\theta_{s}$ is the source size and $\theta_{b}$ is the telescope beam size. 

We calculated the dust optical depth using the same power law assumed in \fbp,
\begin{equation}
\tau_{d}(\nu) = 2 \rm m_{H} (\chi_{dust}) N_{H_{2}} \kappa_{1.3mm} \left( \frac{\nu}{230 \ GHz} \right)^{\beta}
\label{e-dust}
\end{equation}
where \mhyd\ is the mass of a hydrogen atom, \chidust\ is the dust to gas mass ratio, \nhh\ is the column density of \hh\ molecules, \opaco\ is the dust opacity at 1.3~mm (230~GHz), and $\beta$ is the dust spectral index. We assume \opaco~=~0.42 cm$^{2}$ g$^{-1}$, corresponding to an opacity value midway between an MRN distribution \citep{mathis77} and MRN distribution with thin ice mantels, both with no coagulation \citep{ossenkopf94}. We also set $\beta$~=~2, \chidust~=~0.01, and \nhh~=~2.5\sn$^{24}$~\cms. The values for \opaco, $\beta$, \chidust, and \nhh\ used to calculate dust optical depth in this study are consistent with those adopted by \fbp\ to compute \taudustnu\ toward the hot core. We note that the \nhh\ estimate used to compute \taudustnu is 8 times larger than that reported in \citet[][\nhh~=~3.1\sn$^{23}$~\cms]{plume12}, which is derived from C$^{18}$O line observations from the HIFI spectral survey, but is more consistent with \nhh\ measurements based on mm and sub-mm dust continuum observations \citep{favre11, mundy86, genzel89}, which report \nhh~$\gtrsim$~10$^{24}$~\cms. As described by \fbp, when using Eq.~\ref{e-dust}, the higher \nhh\ value clearly produces better agreement between the data and LTE models for other molecules detected toward the hot core. Because \hts\ transitions are detected throughout the entire HIFI spectrum, \taudustnu\ varies approximately between 0.2 ($\nu$~$\approx$~488~GHz) and 2.4 ($\nu$~$\approx$~1900~GHz). When estimating an uncertainty for \taudustnu, we assume a 30\% error for \nhh, so that the \hh\ column density reported in \citet{plume12} lies within 3~$\sigma$ of 2.5\sn$^{24}$~\cms. We do not include an uncertainty for \opaco\ because it does not significantly increase the error in \taudustnu. A 10\% uncertainty in the dust opacity encompasses values for \opaco\ consistent with MRN and MRN with thin ice mantels (no coagulation) at the 3~$\sigma$ level. Such an uncertainty would increase the error in \taudustnu\ by less than 2\%.

To compute upper state column densities, we need an estimate of the source size, which can be obtained from the extremely optically thick \htsi\ lines. In the optically thick limit, the observed peak intensity of a spectral line becomes the product of the beam filling factor and the source function J(T$_{ex}$) attenuated by the dust optical depth, 
\begin{equation}
T_{mb}=\eta_{bf}J(T_{ex})e^{-\tau_{d}(\nu)},
\end{equation}
where the above expression assumes \tex\ is much greater than the background temperature. Because we do not satisfy the condition $h\nu \ll kT_{ex}$ at THz frequencies, the approximation J(T$_{ex}$)~$\approx$~T$_{ex}$ is not valid. We must, therefore, substitute the full expression for the source function. Doing this and solving for T$_{ex}$ we get the following expression,
\begin{equation}
T_{ex}=\frac{h\nu}{k}\left[ln\left(\frac{h\nu}{k}\frac{\eta_{bf} \ e^{-\tau_{d}(\nu)}}{T_{mb}}+1\right)\right]^{-1} .
\label{e-tex}
\end{equation}
We are thus able to estimate the source size by varying it in Eq.~\ref{e-tex} until \tex\ is commensurate with what we expect from the Orion hot core. 

We compute a mean \tex\ for all \htsi\ transitions with \eup~$<$~200~K (10 lines). In addition to being the most optically thick lines in our dataset, these transitions have A$_{ul}$~$\sim$10$^{-3}$~s$^{-1}$ and collision rates of order 10$^{-11}$~cm$^{-3}$s$^{-1}$, yielding critical densities $\sim$~10$^{8}$~\cmc. The density of the hot core is typically estimated to be $\gtrsim$10$^{7}$~\cmc\ \citep{genzel89} and our own non-LTE analysis of the more optically thin \hts\ isotopologues (Sec.~\ref{s-lvg}) requires densities $\gtrsim$10$^{8}$~\cmc\ in order to reproduce the observed emission. Thus, the lowest energy \hts\ transitions should have level populations close to or in LTE, depending on the actual density of the \hts\ emitting gas. Consequently, computed values of T$_{ex}$ for these transitions should be commensurate with the kinetic temperature of the gas in the hot core. Kinetic temperatures toward this region have been measured previously using NH$_{3}$ inversion transitions, which have LTE level populations at hot core densities. Using states with \eup~$\lesssim$~1200~K, previous studies have derived kinetic temperatures of $\sim$160~K toward the hot core \citep{hermsen88, wilson00}. We note that \citet{goddi11} derived kinetic temperatures as high as 490~K in the hot core using NH$_{3}$ inversion lines. This study, however, used transitions with \eup~$\approx$~400 -- 1500~K and is likely sensitive to hotter gas than is predominately probed by lower energy lines. For \ssize\ values in the range 4~--~8\arcsec, we derive a range in mean \tex\ of 300~--~101~K. Source sizes much smaller than 4\arcsec\ result in unreasonably high values for \tex, while sizes larger than 8\arcsec\ yield \tex\ estimates lower than 100~K indicating that the low energy \hts\ transitions are out of LTE. We therefore adopt an intermediate value of 6\arcsec, corresponding to \tex~=~153~K, with an estimated uncertainty of $\pm$~0.7\arcsec\ (3~$\sigma$ thus encompasses the above range in \ssize). 

Table~\ref{t-nup_thick} lists values for $\tau_{iso}$ and \nup(\htsi) for transitions where the optical depth could be explicitly computed and assumes a source size of 6\arcsec. The table lists \nup(\htsi) values assuming both Solar and Orion KL isotopic ratios. Examining Table~\ref{t-nup_thick}, one sees that in instances when the same transition is observed by \htsii\ and \htsiii\ (i.e. 4$_{2,3}$--4$_{1,4}$, 4$_{2,3}$--3$_{1,2}$, and 4$_{4,1}$--4$_{3,2}$), the Orion KL ratios yield upper state columns that are in better agreement with one another. Consequently, we take the isotopic ratios derived by \citet{tercero10} to be more compatible with the observed emission, indicating that the $^{32}$S/$^{33}$S ratio is $\sim$1.7 times smaller than Solar toward the Orion KL hot core. We note that for three transitions (2$_{1,2}$--1$_{0,1}$,  2$_{2,1}$--1$_{1,0}$, and 3$_{3,0}$--2$_{2,1}$) computing values for \htsiii/\htsi\ in order to obtain \tauiso\ resulted in ratios $>$~1 (i.e. the \htsiii\ line is stronger than the corresponding \htsi\ line). There are several reasons why these transitions may have yielded \htsiii/\htsi\ line ratios $>$~1. First, they may be a result of unidentified blends with other transitions. Second, because the line profiles contain several spatial/velocity components, we may have underestimated the contribution of the hot core in the \htsi\ line profiles during the Gaussian fitting process. Conversely, we may have overestimated the hot core contribution in the \htsiii\ profiles for the same reason. And third, it is possible the \htsiii\ lines are still quite optically thick at low excitation. As a result, estimates of $\tau_{iso}$ and \nup\ based on \htsiii/\htsi\ line ratios may be underestimated particularly at low excitation energies. Fortunately, several low energy states are also probed by the more optically thin \htsii\ isotopologue.

\subsection{Rotation Diagram Analysis}
\label{s-lte}

We use the upper state column densities given in Table~\ref{t-nup_thick} to construct rotation diagrams for ortho (upper panel) and para (lower panel) \hts\ in Fig.~\ref{p-rd} assuming Orion KL isotopic ratios. Points were placed on the diagrams by dividing \nup\ for each state by the statistical weight, g$_{u}$. States with multiple \nup\ measurements were averaged together and the uncertainty was propagated by adding the individual errors in quadrature and dividing by the number of measurements. These values were then plotted as a function of energy in units of Kelvin. From these plots, a ``rotation temperature", \trot, and total \htsi\ column density, \ntot, can be derived by performing a linear least squares fit to the points using the following relations,
\begin{equation}
T_{rot}=\frac{1}{m} \;\;\;\;\;  N_{tot}=Q(T)e^{b}
\end{equation}
where m and b are the slope and y intercept of the linear least squares fit, respectively, and Q(T) is the partition function. In the LTE limit, the points should follow a line and \trot\ will equal the kinetic temperature. 

From the figure, we immediately see that the points follow a straight line for both ortho and para \hts, given the uncertainties, over the range in \eup\ where \nup\ values could be computed. If \hts\ is indeed in LTE, the lack of curvature in these diagrams indicates that there is not a strong temperature gradient in the \hts\ emitting gas. Another possibility, however, is that a strong far-IR continuum redistributes the population, particularly at higher energies, to produce the observed rotation diagrams. We investigate this possibility in Sec.~\ref{s-lvg}. Linear least squares fits to the points are straightforward to compute. Fitting the ortho diagram yields \trot~=~141~$\pm$~12~K and \ntot(o-\htsi)~=~5.9~$\pm$~1.3~$\times$~10$^{17}$~\cms, while the para diagram gives \trot~=~133~$\pm$~15~K and \ntot(p-\htsi)~=~2.7~$\pm$~0.7~$\times$~10$^{17}$~\cms. The derived values for \trot, thus, agree to within 1~$\sigma$. We, however, take the ortho rotation temperature as more reliable because upper state column densities are measured over a larger range in \eup\ for ortho \hts. Because ortho and para \hts\ should, in principle, have the same temperature, we fit the para \hts\ rotation diagram again with \trot\ fixed to a value of 141~K. Because the rotation temperatures for ortho and para \hts\ are in close agreement with one another, \ntot(p-\htsi) shifts to a value of 2.4~$\pm$~0.6~$\times$~10$^{17}$~\cms, well within 1~$\sigma$ of the previous value. Adding \ntot(o-\htsi) and \ntot(p-\htsi) gives a total \htsi\ column density of 8.3~$\pm$~1.4~$\times$~10$^{17}$~\cms. 

We emphasize that our derived values for \trot\ and \ntot\ assume \ssize~=~6\arcsec\ and note that our \trot\ estimate agrees well with the mean \tex\ we calculated in Sec.~\ref{s-nup} for this source size. Increasing or decreasing \ssize\ results in a decreasing or increasing \nup\ estimate, respectively, for any given state. The magnitude of this shift depends on the change in source size and the frequency of the transition due to the dependance of telescope beam size on frequency. Because \eup\ is not a strong function of frequency, the net effect of changing the source size on the rotation diagram is a net shift toward higher or lower values for \ntot, with some change in the scatter. Thus, the assumed source size influences \ntot\ but has little affect on \trot. If, however, the highly excited \hts\ emitting gas is significantly more compact than gas traced by low energy lines, the derived values for \nup\ will be shifted up more for higher energies compared to lower ones, resulting in a hotter \trot. The extent to which this is true can only be determined by interferometric maps of \hts, which do not currently exist. 

\subsection{Ortho/Para Ratio of \hts}
\label{s-op}

We are able to estimate the ortho/para ratio of \hts\ in two ways. The first method is simply dividing our measurements for \ntot(o-\htsi) and \ntot(p-\htsi) given in Sec.~\ref{s-lte}, which yields 2.5~$\pm$~0.8 and, of course, assumes LTE and a rotation temperature of 141~K. The second method, which does not assume LTE and is temperature independent, uses the upper state column densities derived in Table~\ref{t-nup_thick} directly. In three instances, we measured \nup\ for ortho and para states with approximately equal upper state energies. The \nup\ ratio between such a pair of states should then reflect the global ortho/para ratio. Table~\ref{t-op} gives ortho/para ratio estimates in the instances where this is possible using the Orion KL isotopic ratios. Based on Table~\ref{t-op}, we compute a mean ortho/para ratio of 1.7~$\pm$~0.8. Both methods, therefore, point to an \hts\ ortho/para ratio smaller than 3, the expected value in thermal equilibrium, but are not statistically different from the equilibrium value at the 3~$\sigma$ level.

\subsection{Non-LTE Analysis: \hts\ Column Density and Abundance}
\label{s-lvg}
We modeled the observed emission using RADEX \citep{vandertak07}, which is a non-LTE code that explicitly solves the equations of statistical equilibrium. RADEX employs the escape probability method to decouple the radiative transfer and statistical equilibrium equations from one another. Because the \hts\ lines have observed widths that far exceed the expected thermal width at typical hot core temperatures (e.g. $\sim$0.5~km/s at \tkin~$\sim$~150~K), we infer that turbulent motions are significant. We therefore use the large velocity gradient (LVG) approximation when running RADEX to model the \hts\ emission toward the hot core. 

We had to estimate collision rates for H$_{2}$~$+$~\hts\ because they have not been measured in the laboratory. We do this in two ways. In the first method, neutral-impact collision rates are scaled in proportion to radiative line strengths so that the sum of all downward rates from each upper state is equal to the base rate of 1.35~$\times$~10$^{-11}$~cm$^{3}$~s$^{-1}$. This base rate value was determined from measured depopulation cross sections reported by \citet{ball99} for the 1$_{1,0}$--1$_{0,1}$ transition for He~$+$~\hts\ collisions interpolated at 10~K and scaled to the reduced mass of the H$_{2}$~$+$~\hts\ system. In the second method, we estimate \hts\ collision rates with ortho and para-\hh\ by scaling existing \hto\ rates from \citet{faure07} such that the para-\hh\ (similar to He as a collision partner) rate for the 1$_{1,0}$--1$_{0,1}$ transition is consistent with the experimental result at low temperature. A constant factor of 0.4 was thus applied to the \hto\ rates to achieve this agreement. We chose \hto\ as an analogue to \hts\ because of its similar molecular structure. We note that, in general, rotational excitation rates are sensitive to details of the interaction potentials. However, these rates are likely a more robust estimate for \hts\ than those scaled from the 1$_{1,0}$--1$_{0,1}$ transition, and provide an example for comparison of how model line intensities might be sensitive to uncertainties in the collision rates. We assume that the H$_{2}$ populations are thermalized at the kinetic temperature. The ortho/para ratio of H$_{2}$ is thus governed by the expression,
\begin{equation}
\frac{o-H_{2}}{p-H_{2}}=9e^{-170.5/T_{kin}}.
\end{equation}
We will refer to the collision rates estimated using the first and second methods described above as CR1 and CR2 rates, respectively, in the text that follows. The CR2 rates, (specifically ortho-H$_{2}$, the dominant form of molecular hydrogen at hot core temperatures), are larger than the CR1 rates by a factor of $\gtrsim$~2. As a result, higher densities are required when using the CR1 rates to achieve the same level of excitation as the CR2 rates. The CR2 rates, however, have a disadvantage in that collision rates for the most highly excited lines in our dataset are not computed. 

We also include a model continuum to investigate if radiation plays a significant role in the excitation of \hts. The continuum model is based directly on observed fluxes of IRc2, the brightest ``IR clump" for $\lambda$~$\lesssim$~10~\um\ and also spatially close to the position of the Orion KL hot core \citep[see e.g.][]{debuizer12, okumura11, greenhill04, gezari98}. Our model continuum is plotted in Fig.~\ref{p-bg_rad}. The observations used to construct the diagram were obtained from a variety of sources \citep{vandishoeck98, lerate06, lonsdale82, dicker09}. The shape and color of each point indicates the reference from which it was obtained. We also measured the continuum at several frequencies throughout the HIFI scan, which are plotted as red triangles. When comparing the HIFI measurements to those obtained with LWS on board ISO (blue circles), we noticed that the continuum level as measured by HIFI was higher by approximately a factor of 4 at frequencies where the two observations overlapped. This difference is likely the result of ISO's larger beam size relative to \herschel. Because our aim is to construct a continuum model that best describes the radiation environment of \hts\ as observed by HIFI, we normalized the ISO LWS and SWS continuum observations (blue circles and squares) so that the LWS continuum matched that of HIFI at $\sim$158~\um. The normalized values are given as black circles and squares, and our final continuum model is plotted as a black dashed line. We also overlay a 65~K blackbody in Fig.~\ref{p-bg_rad} (green, dotted line) normalized to the SWS observations at $\sim$44~\um, which reproduces much of the observed flux in the mid/far-IR and sub-mm. 


Given a radiation field, the input parameters to the RADEX model are the kinetic temperature, \tkin, H$_{2}$ number density, \nht, and total \hts\ column density, \ntot. We can limit the parameter space of \ntot\ using a method originally described by \citet{goldsmith97} and more recently employed by \citet{plume12} to compute column densities for C$^{18}$O toward different Orion KL spatial/velocity components using the same dataset presented here. This method allows us to estimate \ntot(\htsi) by computing a correction factor, CF, to account for the column density not directly probed by HIFI,
\begin{equation}
N_{\rm tot} = CF \times N_{\rm obs}.
\end{equation}
Here, \nobs\ is the column density directly measured by the HIFI survey,
\begin{equation}
N_{\rm obs} = \sum\limits_{i}^{\rm obs} N_{i}
\label{e-nobs}
\end{equation}
where \nsi\ is the column density of a state represented by the index $i$, and the sum is carried out over all states with an observed column density. Adding the \nup\ measurements in Table~\ref{t-nup_thick} (Orion~KL isotopic ratios), we obtain \nobs~=~3.1~$\pm$~0.4~$\times$~10$^{17}$~\cms, where states with more than one measurement for \nup\ were averaged together in the same way as described in Sec.~\ref{s-lte}. 

We estimate the correction factor by running a series of RADEX models over the temperature range 100 -- 400~K for H$_{2}$ densities of 10$^{7}$, 10$^{8}$, and 10$^{9}$~\cmc, i.e. physical conditions expected within the hot core. All models assume a column density of 5~$\times$~10$^{16}$~\cms\ to avoid any problems with convergence associated with very optically thick low energy lines. For each model realization, we compute CF by dividing the assumed total column density (\ntot~=~5~$\times$~10$^{16}$~\cms) by the \nobs\ predicted by RADEX. Fig.~\ref{cf} plots values for CF as a function of temperature using both CR1 (black) and CR2 (red) collision rates. For comparison, we also plot correction factors that assume LTE (blue, dot-dashed line) in Fig.~\ref{cf}. These CF values are computed using the equation,
\begin{equation}
CF= \frac{Q(T_{\rm kin})}{\sum\limits_{i}^{\rm obs} g_{i}e^{-E_{i}/kT_{\rm kin}}},
\end{equation} 
where $Q(T_{\rm kin})$ is the partition function, and $g_{i}$ and $E_{i}$ are the statistical weight and excitation energy, respectively, of a state represented by the index $i$. Just as in Eq.~\ref{e-nobs}, the sum in the denominator is carried out over all states with an observed column density. At \tkin~=~140~K, approximately equal to our derived \trot, the LTE correction factor is commensurate with CF values derived from RADEX models with \nht~$\ge$~10$^{8}$~\cmc. RADEX models which set \nht~=~10$^{7}$~\cmc\ and assume the CR1 collision rates produce correction factors that are significantly higher than the other RADEX models. The reason for this deviation is that using these rates (which are smaller than the CR2 rates), a density of \nht~=~10$^{7}$~\cmc\ is not high enough to excite a significant fraction of the \hts\ population above J=1, which is not probed by the HIFI scan. The assumption of LTE, on the other hand, produces correction factors that increase with temperature because more population is pushed into highly excited states where we do not measure \nup. The fact that the RADEX models either produce correction factors that do not increase with \tkin\ or CF curves that are shallower than the one corresponding to LTE indicates RADEX predicts sub-thermal excitation even when \nht~=~10$^{9}$~\cmc. This method of estimating \ntot(\htsi) thus differs from our rotation diagram approach in that it does not require us to assume LTE and allows us to estimate \ntot(\htsi) without a corresponding temperature estimate because CF is not a strong function of \tkin\ according RADEX.

Fitting the observed \htsii\ and \htsiii\ emission to a grid of RADEX models indicates that the \hts\ emission originates from extremely dense gas (\nht~$>$~10$^{8}$~\cmc; see text below and Figs.~\ref{p-chisq_orion} and \ref{p-chisq_ir100sc30}). We therefore take the correction factors for \nht~$\geq$~10$^{8}$~\cmc\ as most representative of the \hts\ emitting gas in the Orion KL hot core. For these high densities, our derived values for CF have a maximum range of 2.6--3.5 (excluding LTE), thus we have detected approximately 29 -- 38\% of the total population with the HIFI scan. We adopt a correction factor of 3.05~$\pm$~0.45, where the uncertainty encompasses the entire range in CF, which yields \ntot(\htsi)~=~9.5~$\pm$~1.9~$\times$~10$^{17}$~\cms. Adopting an H$_{2}$ column density of 3.1~$\times$~10$^{23}$~\cms\ toward the Orion hot core \citep{plume12}, we obtain an \htsi\ abundance of 3.1~$\times$~10$^{-6}$. The \nhh\ value derived by \citet{plume12}, however, assumes a hot core source size of 10\arcsec. We, on the other hand, derive a 6\arcsec\ source size for the \hts\ emitting gas in the Orion~KL hot core. Using Eq.~\ref{e-bff} and setting \bsize~=~20\arcsec, (the \herschel\ beam size at $\nu$~$\sim$~1~THz), and the fact that \nup\ is indirectly proportional to \beamff\ (Eq.~\ref{e-nup_thick}), we estimate that adjusting \ssize\ to 6\arcsec\ would increase the \hh\ column density reported by \citet{plume12} by a factor $\sim$~2.5, resulting in an \htsi\ abundance of 1.2~$\times$~10$^{-6}$. The implicit assumption here is that CO and \hts\ are emitting from the same region within the hot core. We thus report an \htsi\ abundance of 3.1~$\pm$~$^{1.1}_{1.9}$~$\times$~10$^{-6}$, where the positive uncertainty is obtained by propagating the uncertainties in \ntot(\htsi) and \nhh\ from \citet{plume12} and the negative uncertainty is set so that an abundance of 1.2~$\times$~10$^{-6}$ lies within 1~$\sigma$. Assuming solar metallicity, our abundance corresponds to approximately 12\% of the available sulfur in \hts.

\subsection{Non-LTE Analysis: Kinetic Temperature and \hh\ Density}

Having constrained \ntot(\htsi), we now determine which values of \tkin\ and \nht\ best reproduce the observed emission. In order to do this, we constructed a grid of RADEX models that varied \tkin\ and \nht. The grid covered the following ranges in parameter space: \nht~=~10$^{7-12}$~\cmc, evaluated in logarithmic steps of 0.1, and \tkin~=~40 -- 800~K, evaluated in linear steps of 20~K. Grids were computed using both the CR1 and CR2 rates. In order to compare our data to the grid, we carried out reduced chi squared, \redchisq, goodness of fit calculations using \tpeak\ as the fitted parameter. The sources of uncertainty used in the \redchisq\ calculations are described in the Appendix.

We noticed when \ntot\ approached 10$^{18}$~\cms, commensurate with our derived value for \ntot(\htsi), that RADEX either failed to properly converge or the code produced a segmentation fault for some regions in the parameter space. Furthermore, using such a high column density produced optical depths in excess of 100 in some low energy lines for certain combinations of \tkin\ and \nht. Predicted line intensities for such transitions can not be trusted when using RADEX \citep{vandertak07}. As a result, the \redchisq\ was evaluated using only the \htsii\ and \htsiii\ lines, which have lower optical depths than the main isotopologue. We thus had 32 data values with which to compare to our models\footnote{We exclude the \htsii\ and \htsiii\ 6$_{0,6}$--5$_{1,5}$ lines from our fit because of self-blending with the 6$_{1,6}$--5$_{0,5}$ transition.}. There are two fitted parameters: \tkin\ and \nht, yielding 30 degrees of freedom. We set \ssize\ and \ntot(\htsi) to our derived values of 6\arcsec\ and 9.5$\times$~10$^{17}$~\cms, respectively. Each grid point, then, included two model realizations, one for \htsii\ and \htsiii, with \ntot\ values scaled according to Orion KL isotopic ratios. That is, we set \ntot(\htsiii)~=~4.8\sn$^{16}$~\cms\ and \ntot(\htsii)~=~1.3\sn$^{16}$~\cms. We also set \dv\ equal to 8.6~km/s, the average measured line width in our dataset for the hot core. 


The upper and lower panels of Fig.~\ref{p-chisq_orion} are \redchisq\ contour plots for model grids using the CR1 and CR2 rates, respectively. The contours correspond to p values of 0.317 (\redchisq~=~1.1), 0.046 (\redchisq~=~1.5), and 0.003 (\redchisq~=~1.9), representing 1, 2, and 3~$\sigma$ confidence intervals, respectively. Here, ``p value" is the probability that a random set of data points drawn from distributions represented by our uncertainties will result in a \redchisq\ statistic equal to or greater than a given value \citep[see Sec. 4.4 of][]{bevington03}. Models lying outside the largest contour in Fig.~\ref{p-chisq_orion} are thus statistically inconsistent with the data. We see from the figure that, for both sets of collision rates, high density solutions are necessary to fit the data. We constrain \nht\ to be $\gtrsim$~3\sn$^{8}$~\cmc\ based on the CR2 rates, which require lower density solutions relative to the CR1 rates by a factor of $\sim$~10 to achieve the same goodness of fit. This is a result of the fact that the CR2 rates are larger than the CR1 rates. Combining the ranges in \tkin\ covered by the 1~$\sigma$ contours for the CR1 and CR2 grids, we constrain \tkin~=~130$\pm^{33}_{15}$~K. 

The best fit solutions produce good agreement over the entire range in upper state energy over which \htsii\ and \htsiii\ are detected. Fig.~\ref{p-he34} compares the observed \htsiii\ lines to a RADEX model which sets \tkin~=~130~K and \nht~=~7.0\sn$^{9}$~\cmc, a solution within the 1~$\sigma$ confidence interval assuming CR2 collision rates and the observed continuum (Fig.~\ref{p-chisq_orion}, lower panel). The large bottom panel plots the ratio of predicted to observed \tpeak\ as a function of \eup\ as blue points. Fig.~\ref{p-he34} also plots a sample of six \htsiii\ lines, which spans the range in \eup\ over which \htsiii\ emission is detected, arranged so that \eup\ increases from the lower left panel to the upper right. The same model represented as blue points in the \tpeak\ vs. \eup\ panel is overlaid as a blue solid line in each small panel. Fig.~\ref{p-he33} is an analogous plot which compares the observed \htsii\ lines to the same model plotted in Fig.~\ref{p-he34} with \ntot(\htsii) scaled according to Orion~KL isotopic ratios. As in Fig.~\ref{p-he34}, the model is represented in blue. Taking this model as representative of the best fit solutions, Figs~\ref{p-he34} and \ref{p-he33} show that there is no systematic trend in the model residuals as a function of \eup\ for both \htsii\ and \htsiii.

The best fit solutions to the isotopic emission, however, under predict the most highly excited \htsi\ lines. Fig.~\ref{p-he_orion} is a plot of eight highly excited \htsi\ transitions (black) with four RADEX models overlaid. The models assume our derived value of \ntot(\htsi)~=~9.5~$\times$~10$^{17}$~\cms\ and use CR1 rates because the CR2 values do not include collisions into these states. Different curves correspond to models with \nht~=~1.5\sn$^{10}$ and 1.5\sn$^{11}$~\cmc\ (blue and red lines) and \tkin~=~140 and 300~K (dashed and solid lines). We know from our analysis in Sec.~\ref{s-lte} that an LTE solution with \trot~$\approx$~140~K reproduces the observed emission. Above the critical density for these transitions we, therefore, expect RADEX to predict line intensities comparable to what is observed for \tkin~$\approx$~140~K. Fig.~\ref{p-he_orion} shows that this occurs at densities $\gtrsim$~10$^{11}$~\cmc\ (dashed red line), though the emission is still somewhat under predicted. Assuming a distance of 414~pc \citep{menten07} and that 74\% of the mass is in hydrogen, a spherical clump with a 6\arcsec\ diameter and uniform \hh\ density $>$~10$^{11}$~\cmc\ corresponds to a total clump mass $\gtrsim$~6100~M$_{\sun}$. Such a large value is not consistent with mm observation, which report total clump masses $\lesssim$~40~M$_{\sun}$ for the Orion KL hot core \citep[e.g.][]{favre11,wright85, wright92}. Even if our clump mass lower limit is overestimated by an order of magnitude because of uncertainties in the collision rates, we still predict values a factor of $\sim$15 larger than what mm-observations derive. We therefore conclude that LTE solutions require unrealistically high densities. Alternatively, Fig.~\ref{p-he_orion} also shows that the highly excited \htsi\ emission is well fit at lower densities if \tkin~$\approx$~300~K (solid blue line). Such solutions, however, are inconsistent with the isotopic emission. In other words, this range in parameter space (i.e. \nht~$\gtrsim$~10$^{10}$~\cmc, \tkin~$\gtrsim$~300~K) exists outside the reduced \chisq\ contour of 1.9 in Fig.~\ref{p-chisq_orion}. 

\subsection{Non-LTE Analysis: Radiative Excitation}

It is clear that unrealistically high densities are required to reproduce the observed \hts\ emission over the entire range in excitation. As mentioned above, our RADEX models incorporate the observed continuum toward IRc2. However, it is possible that this radiation field is an under estimate of that seen by \hts, perhaps as a result of optical depth or geometrical dilution.  Increasing the far-IR/sub-mm continuum in order to pump \hts\ via the same transitions observed in our dataset tends only to push the predicted intensities of observed lines into absorption. However, if there is a source of luminosity within the hot core, deeply embedded hot dust emitting heavily in the mid-IR and the short wavelength end of the far-IR, where the hot core continuum peaks (i.e. $\lambda$~$\lesssim$~100~\um), may be hidden because the continuum is optically thick. The dust optical depth estimates presented by \fbp\ would certainly imply an optically thick continuum for $\lambda$~$\lesssim$~100~\um\ (see Sec.~\ref{s-nup}). Using Eq.~\ref{e-dust} and assuming the same values as Sec.~\ref{s-nup}, we estimate \taudust(100~\um)~$\approx$~6. \citet{genzel89}, moreover, argue that the far-IR continuum may be optically thick based on the shape of the spectral energy distribution between 400~\um\ and 3~mm. We therefore conclude that the continuum is likely quite optically thick for $\lambda$~$\lesssim$~100~\um\ toward the Orion KL hot core.

If \hts\ is indeed probing heavily embedded gas near a hidden self-luminous source, the continuum seen by \hts\ may be enhanced relative to what is observed especially where the continuum is most optically thick. Consequently, we enhanced the background continuum for $\lambda$~$<$~100~\um\ in order to determine if this resulted in better agreement between the data and models. Fig.~\ref{p-chisq_ir100sc30} is a reduced \chisq\ contour plot produced using the same methodology as Fig.~\ref{p-chisq_orion}, except the assumed background continuum is enhanced  by a factor of 8 for $\lambda$~$<$~100~\um. We also note that when defining a radiation field within the RADEX program, the user can specify a dilution factor. For the observed continuum, we set the dilution factor to 0.25 or 1.0 for wavelengths less than or greater than 45~\um, respectively. For the enhanced continuum, we set the dilution factor to 1.0 for all wavelengths, indicating the dust is co-spatial with the \hts\ emitting gas. From Fig.~\ref{p-chisq_ir100sc30}, we see that, as a result of the enhanced radiation field, the 3~$\sigma$ lower limit on \nht\ is shifted lower by approximately a factor of 3 for both the CR1 and CR2 rates. Using these contour plots in the same way as before, we constrain \nht~$\gtrsim$~9\sn$^{7}$~\cmc\ and \tkin~=~120$\pm^{13}_{10}$~K. 

As with the observed continuum, the best fit solutions using the enhanced radiation field reproduce the observed \htsii\ and \htsiii\ emission well over all excitation energies. Predicted to observed \tpeak\ ratios using a model with \tkin~=~120~K and \nht~=~3.0\sn$^{9}$~\cmc\ and assuming the enhanced continuum are overlaid in Fig.~\ref{p-he34} as red points in the large bottom panel and as dashed red lines in the six smaller upper panels. This solution lies within the 1~$\sigma$ contour in Fig.~\ref{p-chisq_ir100sc30} assuming CR2 collision rates (lower panel). From the figure, we see good agreement between the data and model at all excitation energies. The enhanced continuum, furthermore, produces better agreement with the most highly excited \htsi\ lines. Fig.~\ref{p-he_sc} plots the same highly excited \htsi\ lines given in Fig.~\ref{p-he_orion}. The overlaid model, which is plotted in red, corresponds to \tkin~=~120~K and \nht~=~3.0\sn$^{9}$~\cmc\ using the CR1 rates. This is the same model used in Fig.~\ref{p-he34} and lies within the 2~$\sigma$ contour of the enhanced continuum CR1 grid (Fig.~\ref{p-chisq_ir100sc30}, upper panel). The plot shows good agreement between these highly excited transitions and the model, indicating that an intense far-IR ($\lambda$~$<$~100~\um) radiation field can excite these states for densities and temperatures consistent with the isotopic emission.

The mechanism by which \hts\ is pumped is illustrated in Fig.~\ref{p-gd_irpump}. The plot shows the same energy level diagram given in Fig.~\ref{p-gd}, except the lines connecting the levels are now transitions with $\lambda$~$<$~100~\um\ and $\mu^{2}$S~$>$~0.01. From the plot, we see that there are a litany of $\Delta$J~=~1 transitions for J~$\gtrsim$~3 through which the background radiation field can excite \hts. However, the $\lambda$~$<$~100~\um\ continuum is unable to pump states below J~=~3~--~4 because these transitions occur at longer wavelengths (probed by HIFI), where the continuum is more optically thin.  High densities, therefore, are still required to populate states up to J~=~3~--~4 in order for \hts\ to be pumped to more highly excited states. This is somewhat different from traditional pumping as seen, for example, in OH, where the ground state transitions couple directly to the intense radiation field which pumps to higher levels. In the case of \hts, the pumping requires excitation via collisions to higher states in order for radiation to redistribute the population. We note that although we nominally increased the continuum for $\lambda$~$<$~100~\um, the transitions plotted in Fig.~\ref{p-gd_irpump} occur at wavelengths in the range 39~--~100~\um. It is thus the short wavelength end of the far-IR that is responsible for the \hts\ pumping. 

Fundamental vibrational transitions for \hts\ occur at wavelengths $\lesssim$~8.5\um. In this wavelength region, the observed radiation field is weaker than the continuum responsible for pumping \hts\ rotation transitions (39~--~100\um) by over an order of magnitude. Nevertheless, we searched for the presence of vibrationally excited \hts\ in the HIFI scan. Vibrational transitions were obtained from the HITRAN database. Although these measurements did not include pure rotation transitions within an excited vibrational state, (and as of yet have not been measured in the laboratory), we computed approximate frequencies for several rotation lines with J~$<$~4 within the $\nu_{1}$, $\nu_{2}$, and $\nu_{3}$~=~1 states by taking upper state energy differences.  Given that \eup\ values for these states are reported to a precision of 1 part in 10$^4$, the frequencies should be accurate to within a few km/s in the HIFI scan. We searched for emission near these frequencies by plotting the data with the full band model overlaid to identify line blends. We did not detect unidentified lines at these frequencies that could be attributed to vibrationally excited \hts, but note that many of these transitions were in spectral regions of high line density making it difficult to search for such features.

We also investigated the significance of vibrational excitation in our models by computing HITRAN versions of the CR1 and CR2 collision rates. We produced a coarse model grid with \ntot~=~10$^{17}$~\cms\ and varied \nht\ and \tkin\ within ranges 10$^{7}$~--~10$^{10}$~\cmc, evaluated in logarithmic steps of 1, and 100~--~400~K, evaluated in 100~K steps, respectively. For each grid point, we produced a model with and without vibrational excitation and compared the predicted line fluxes. In most instances, these values agreed to within 10\%, provided a given state was sufficiently populated to produce detectable emission. This agreement held when using both the observed and enhanced continuum models. We thus conclude the presence of vibrational excitation does not significantly affect our results, assuming the two radiation fields discussed in this work. The caveat here is that if \hts\ is probing gas near a hidden source of luminosity, the shape of the continuum seen by \hts\ may be more strongly peaked in the mid-IR possibly producing higher levels of vibrational excitation.   

\subsection{D/H Ratio Upper Limit}
\label{s-hds}

We do not detect HDS emission toward Orion KL. In order to compute an upper limit for the column density of HDS, we looked at six line free regions in the HIFI scan where emissive low energy HDS lines should be present. The transitions we used are given in Table~\ref{t-hds}. We used the XCLASS program\footnote{http://www.astro.uni- koeln.de/projects/schilke/XCLASS} to compute model spectra, which assumes LTE level populations. We also assumed a source size of 6\arcsec, \dv=8.6~km/s, and \trot~=~141~K, in agreement with our rotation diagram analysis (Sec.~\ref{s-lte}). We then increased the column density until the peak line intensities exceeded 3~$\times$ the local RMS in all lines. This procedure yields an upper limit on the column density of \ntot(HDS)~$<$~4.7\sn$^{15}$~\cms. Using the \hts\ column density derived from our non-LTE analysis (Sec.~\ref{s-lvg}), this corresponds to a D/H ratio $<$~4.9\sn$^{-3}$. 

\section{Discussion: Origin of \hts\ emission}
\label{s-disc}

Our results indicate that \hts\ is a tracer of extremely dense gas toward the Orion KL hot core and that the far-IR background continuum plays a significant role in the excitation of this molecule especially for the most highly excited transitions in our dataset. Moreover, the observed far-IR  continuum ($\lambda$~$<$~100~\um) toward IRc2 is not sufficient to reproduce the excitation that we detect. These results point to an \hts\ origin in heavily embedded, dense gas close to a hidden source of luminosity that heats nearby dust which cannot be directly observed because the continuum is optically thick in the mid- and short-wavelength-far-IR. Such an object (or objects) may be the ultimate source of Orion KL's high luminosity \citep[$\sim$10$^{5}$~L$_{\sun}$;][]{wynn84} and thus harbor an intense far-IR radiation field responsible for the highly excited \hts\ transitions we observe toward Orion KL.

Finding evidence for the presence of hidden self-luminous sources toward Orion KL is difficult not only because of the high IR optical depth, but also because of its elaborate physical structure. Near to mid-infrared maps reveal the presence of many IR clumps (``IRc" sources), only a small fraction of which may be self luminous \citep[see e.g.][]{rieke73, downes81, lonsdale82, werner83, gezari98, greenhill04, shuping04, robberto05, okumura11, debuizer12}. The situation is further complicated by the presence of radio source I \citep{churchwell87}, a supposed heavily embedded massive protostar, which may be externally heating nearby clumps. \citet{okumura11}, for example, studied the interaction between radio source I and IRc2, the brightest clump in the mid-infrared and once thought to be the source of Orion KL's high luminosity. More recent observations, however, have ruled this out \citep{dougados93, gezari98, shuping04}. In fitting the mid-IR SED of IRc2, \citet{okumura11} find that they can not fit the data with a single blackbody. Instead they require two blackbody components with temperatures of 150 and 400~K, the hotter component supplying most of the flux for $\lambda$~$\lesssim$~12~\um. The origin of the hotter component, they argue, is scattered radiation from radio source I, which lies approximately 1\arcsec\ away from IRc2, and the source toward which they detect a prominent 7.8~/~12.4~\um\ color temperature peak. Evidence of this hotter component can be seen in Fig.~\ref{p-bg_rad} as excess emission for $\lambda$~$\lesssim$~10~\um. Focusing on slightly longer wavelengths, \citet{debuizer12} present mid-IR SOFIA maps obtained with the FORCAST camera which show 19.7~/~31.5~\um\ and 31.5~/~37.5~\um\ color temperature peaks toward IRc4 indicating that it may also be self-luminous. 

Previous observations of highly excited molecular lines in the mm and sub-mm have confirmed the influence of an intense far-IR continuum within the Orion KL hot core. \citet{hermsen88}, for example, present both metastable and non-metastable inversion transitions of \amm\ spanning a large range in excitation energy ($\sim$~1200~K). Comparing their observations to non-LTE models which specify \nht, \tkin, the dust temperature, \tdust, and the dust optical depth at 50~\um, \taudust(50~\um), they find a best fit solution using \nht~=~10$^{7}$~\cmc,  \tkin~=~150~K, \tdust~=~200~K, and \taudust(50~\um)~=~10. The strong far-IR radiation field produced by the hot, optically thick dust, they argue, is necessary to explain the highly excited (\eup~$>$~700~K) non-metastable \amm\ lines. Using a similar methodology, \citet{jacq90} analyzed several HDO transitions spanning a range in \eup\ of approximately 50 to 840~K toward the Orion KL hot core. Their best fit model sets \nht~=~10$^{7}$~\cmc, \tkin~=~\tdust~=~200~K, and \taudust(50~\um)~=~5, again resulting in a strong far-IR background continuum. As a consequence, this model predicts that far-IR pumping is the dominant excitation mechanism for HDO lines with \eup~$\gtrsim$~150~K toward the Orion KL hot core. \citet{goldsmith83}, furthermore, require a strong mid/far-IR radiation field ($\lambda$~$\leq$~44\um) in order to explain their observations of vibrationally excited \mc\ and \hctn. Based on their computed \mc\ vibrational excitation temperature and a scaling law from \citet{scoville76}, they infer dust temperatures $\gtrsim$~260~K for $\theta$~$<$~1\arcsec.

It is clear that the far-IR radiation field plays a significant role in the excitation not only of \hts\ but several other molecular species within the Orion KL hot core. An enhancement in the far-IR continuum can be explained by the presence of hot dust. Assuming that the \hts\ emission originates from embedded hot dust with a temperature of $\sim$200~K as opposed to 65~K, which is more in line with our observed radiation field (Fig.~\ref{p-bg_rad}), we would expect approximately a factor of 9 enhancement at $\sim$~85~\um\ assuming blackbody emission. It is possible that the CR1 collision rates are significantly underestimated for highly excited \hts\ emission, creating a need for a stronger far-IR field. Such a suspicion is reasonable given that the CR1 rates are smaller than the CR2 rates. Furthermore, the presence of both density and temperature gradients may also reduce the need for such a strong enhancement in the far-IR continuum because extremely compact, hot regions may contribute to the \hts\ excitation at high energies. Such an investigation is beyond the scope of this study because it requires more detailed radiative transfer calculations that would involve modeling the physical structure of the Orion KL hot core. Nevertheless, the most highly excited \hts\ lines either require an exceedingly high density or an enhanced continuum. As argued above, the second scenario is more likely. 

Our \emph{Herschel}/HIFI observations contain little spatial information, the true origin of the \hts\ emission therefore cannot be unambiguously determined from our dataset alone. Two pointings, however, were obtained toward Orion~KL in bands~6 and 7 ($\nu$~$>$~1430~GHz) because of the relatively small beam size at these wavelengths ($\lesssim$~15\arcsec). The first pointing, toward the hot core, is given in Sec.~\ref{s-obs} and is almost coincident ($<$~1\arcsec) with radio source I. The second pointing is positioned toward the compact ridge at coordinates $\alpha_{J2000} = 5^h35^m14.1^s$ and $\delta_{J2000} = -5^{\circ}22'36.5''$, and is located 8\arcsec\ SW of the hot core pointing much closer to IRc4 ($\sim$2\arcsec\ away). We also note that the main pointing for bands~1 -- 5 lies midway between the hot core and compact ridge positions. The three pointings, thus, make a NE-SW line across the KL region.  Fig.~\ref{p-tpoint} plots a sample of six \hts\ lines that lie in bands 6 and 7 for the hot core (black) and compact ridge (red) pointings. From the plot, we immediately see that the \hts\ emission is stronger in the hot core pointing relative to the compact ridge, indicating that the \hts\ emission is compact and clumpy and that the majority of  the emission likely originates from a region closer to IRc2/radio source I as opposed to IRc4, the putative self luminous source reported by \citet{debuizer12}. 

Our derived \hts\ abundance is at least two orders of magnitude larger than those measured toward other massive hot cores, which typically have abundances between 10$^{-9}$ and 10$^{-8}$ \citep{hatchell98, vandertak03}. If \hts\ does indeed originate from grain surfaces as has been suggested by \citet{charnley97}, such a large abundance may be the result of increased evaporation from dust grains. Such an interpretation has been invoked for \hto, a structurally similar molecule to \hts, toward the hot core. Using the Orion~KL HIFI survey, \citet{neill13} derive an extremely large \hto\ abundance of 6.5\sn$^{-4}$ and a high water D/H ratio of 3.0\sn$^{-3}$, which they suggest are a result of ongoing evaporation from dust grains. Our measurement of an \hts\ ortho/para ratio $<$~3, though not at a 3~$\sigma$ level, also points to \hts\ formation at cold temperatures possibly on grain surfaces during an earlier, colder stage.  There is, however, no direct observational evidence to support an \hts\ origin on grain surfaces. Near-IR spectroscopic observations of high and low mass protostars have failed to detect \hts\ or any other sulfur bearing species except, possibly, OCS in the solid phase \citep{gibb04, boogert00, smith91, palumbo95, palumbo97}. \citet{goicoechea06}, moreover, find evidence for low depletion levels of sulfur in the Horsehead PDR.  Comparing their measured CS and HCS$^{+}$ abundances to photochemical models, they require a total gas phase sulfur abundance that is $\gtrsim$~25\% the solar elemental abundance of sulfur. Given that our derived \hts\ abundance implies $\lesssim$~12\% of the available sulfur is in the form of \hts, an \hts\ origin on grain surfaces may not necessarily be inconsistent with the Goicoechea and near-IR studies. All currently explored \hts\ gas phase formation routes are highly endothermic, with some key reactions having activation energies $\sim$10$^{4}$~K \citep[see e.g.][and references therein]{desforets93, mitchell84}. Gas phase formation of \hts\ is, therefore, possible in shocks but is difficult to explain at hot core densities and temperatures. Given, however, that high temperature sulfur chemistry is still not well understood, it is possible that \hts\ may have formed in the gas phase by some, as of yet, unexplored route. 

\section{Conclusions}
\label{s-con}

We have analyzed the \hts\ emission toward the Orion KL hot core. In total we detect 52, 24, and 8 unblended or partially blended lines from \htsi, \htsiii, and \htsii, respectively, spanning a range in \eup\ of 55~--~1233~K. Analysis of the extremely optically thick low energy \htsi\ lines indicates that the \hts\ emitting gas is compact (\ssize~$\sim$~6\arcsec). Measured line intensities of the same transition from different isotopologues allowed us to compute upper state column densities for individual states. Using these \nup\ measurements, we constructed rotation diagrams for ortho and para \hts\ and derived \trot~=~141~$\pm$~12~K and \ntot(\htsi)~=~8.3~$\pm$~1.4\sn$^{17}$~\cms\ (ortho + para \hts). We measured the \hts\ ortho/para ratio using two different methods which yield values of 2.5~$\pm$~0.8 and 1.7~$\pm$~0.8, both of which suggest a ratio less than the equilibrium value of 3 but are not statistically inconsistent with thermal equilibrium given the uncertainties. Although we do not detect HDS, we derive a D/H ratio upper limit of $<$~4.9\sn$^{-3}$.

We also modeled the \hts\ emission using the RADEX non-LTE code assuming two different sets of estimated collision rates. We derived a value for the total column density, \ntot(\htsi)~=~9.5~$\pm$~1.9\sn$^{17}$~\cms, by computing a correction factor to account for the \hts\ column not probed by HIFI. Using this column density, we constrain \nht~$\gtrsim$~9\sn$^{7}$~\cmc\ and \tkin~=~120~$\pm$~$^{13}_{10}$~K by comparing a grid of RADEX models to the \htsii\ and \htsiii\ emission. These constraints require that the observed continuum be enhanced by a factor of 8 for $\lambda$~$<$~100~\um. Enhancing the background continuum also produces good agreement between the data and models for the most highly excited \htsi\ lines, which are populated primarily by pumping from the short wavelength end of the far-IR ($\lambda$~$\approx$~39 -- 100~\um), for temperatures and densities consistent with the rarer isotopic emission. We conclude that the \hts\ emitting gas must be tracing markedly dense heavily irradiated gas toward the Orion KL hot core. These conditions point to an \hts\ origin in heavily embedded material in close proximity to a hidden source of luminosity. The source of this luminosity remains unclear. 

\acknowledgments
We thank the anonymous referee who provided comments that improved this manuscript. HIFI has been designed and built by a consortium of institutes and university departments from across Europe, Canada and the United States under the leadership of SRON Netherlands Institute for Space Research, Groningen, The Netherlands and with major contributions from Germany, France and the US. Consortium members are: Canada: CSA, U.Waterloo; France: CESR, LAB, LERMA, IRAM; Germany: KOSMA, MPIfR, MPS; Ireland, NUI Maynooth; Italy: ASI, IFSI-INAF, Osservatorio Astrofisico di Arcetri-INAF; Netherlands: SRON, TUD; Poland: CAMK, CBK; Spain: Observatorio Astron\'{o}mico Nacional (IGN), Centro de Astrobiolog\'{i}a (CSIC-INTA). Sweden: Chalmers University of Technology - MC2, RSS \& GARD; Onsala Space Observatory; Swedish National Space Board, Stockholm University - Stockholm Observatory; Switzerland: ETH Zurich, FHNW; USA: Caltech, JPL, NHSC. 
HIPE is a joint development by the Herschel Science Ground Segment Consortium, consisting of ESA, the NASA Herschel Science Center, and the HIFI, PACS and SPIRE consortia. Support for this work was provided by NASA through an award issued by JPL/Caltech.

\appendix
\section{Uncertainties}
\label{s-unc}

All uncertainties were propagated using the standard ``error propagation equation" \citep[][their Eq.~3.13]{bevington03} unless otherwise stated. The errors reported in Tables~\ref{t-hc}--\ref{t-er} for \vlsr, \dv, \tpeak, and \tint\ are those computed by the CLASS fitting algorithm. Given the spectral resolution of HIFI is $\ge$~0.2~km/s, we assume a minimum error for \vlsr\ and \dv\ of 0.1~km/s and a minimum error for \tint\ of 0.1~K km/s, even if CLASS reports smaller uncertainties for these parameters. The uncertainties listed for \tpeak\ are RMS measurements in the local baseline also calculated using CLASS. 

When propagating errors involving either \tpeak\ or \tint, we include a 10\% calibration error and a pointing uncertainty. The pointing uncertainty is estimated using the following relation,
\begin{equation}
\Delta_{p}=1-exp\left[-~\frac{\sigma_{p}^{2}}{2(\theta_{b}/2.355)^{2}} \right].
\end{equation}
Here, $\Delta_{p}$ is the percent error in either \tpeak\ or \tint\ introduced by the telescope pointing error, $\sigma_{p}$, which we assume to be 2\arcsec\ \citep{pilbratt10}. In order to account for the calibration and pointing errors, we added the CLASS, calibration, and pointing uncertainties in quadrature before performing any further calculations involving \tpeak\ or \tint. We also estimate an uncertainty of 0.7\arcsec\ in our derived source size and a 30\% error in the \nhh\ estimate used to compute \taudustnu\ (see Sec.~\ref{s-nup}), both of which we include as sources of error in our computations of \nup. The source size and \nhh\ uncertainties were also added in quadrature with the RMS, calibration, and pointing errors when computing \redchisq\ values for the model grids. 

\clearpage

\begin{figure}
\plotone{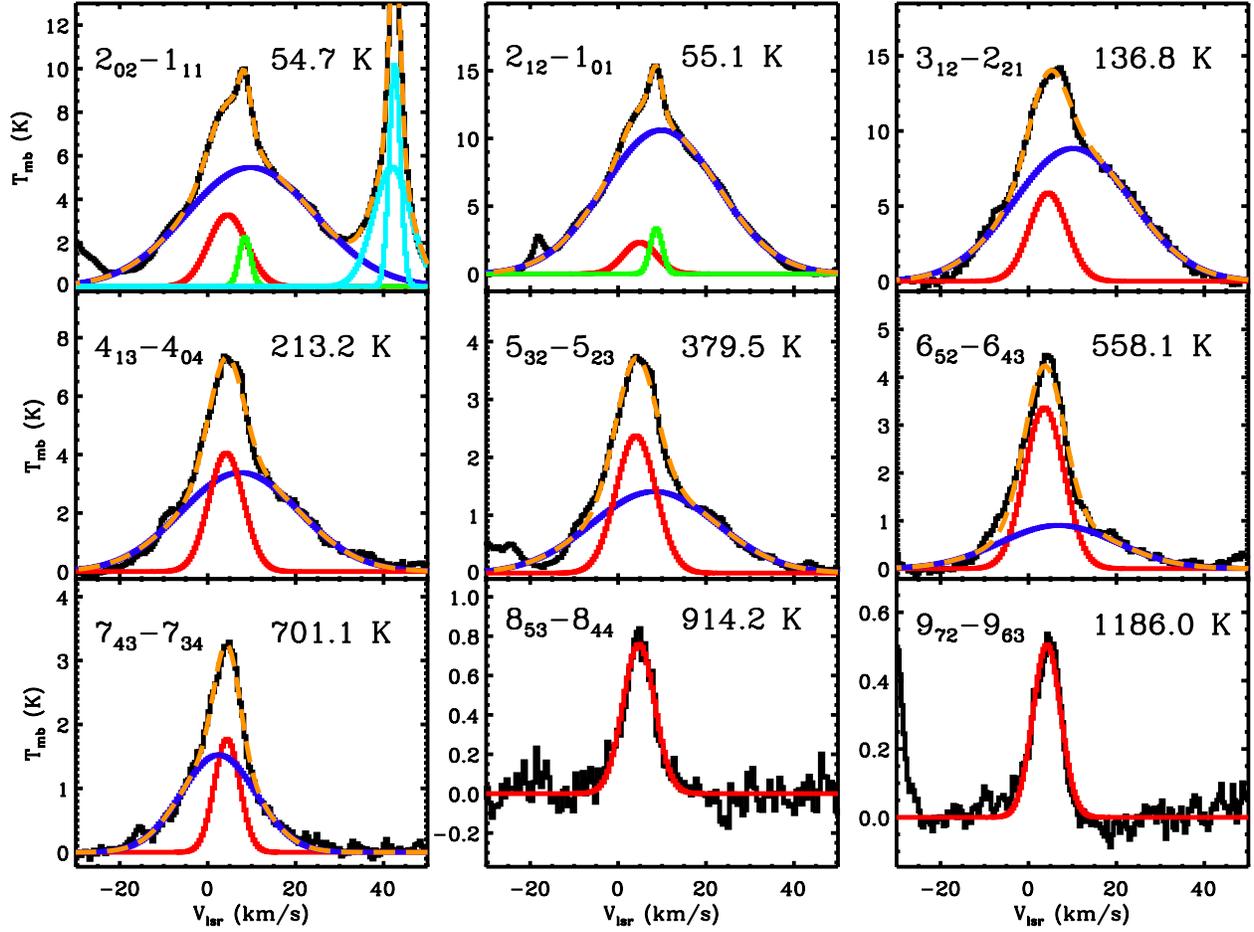}
\caption{A sample of nine \htsi\ lines observed in the HIFI scan with Gaussian fits to the outflow, extended ridge, and hot core spatial/velocity components overlaid in blue, green, and red, respectively. The data are plotted in black and uniformly smoothed to a resolution of $\sim$0.5 km/s. Blending lines that were also fit are plotted in cyan. The sum of all components is plotted as an orange dashed line. The upper state energy of the transition is given in each panel and increases from the upper left panel to the lower right. \label{p-ls}}
\end{figure}

\clearpage

\begin{figure}
\plotone{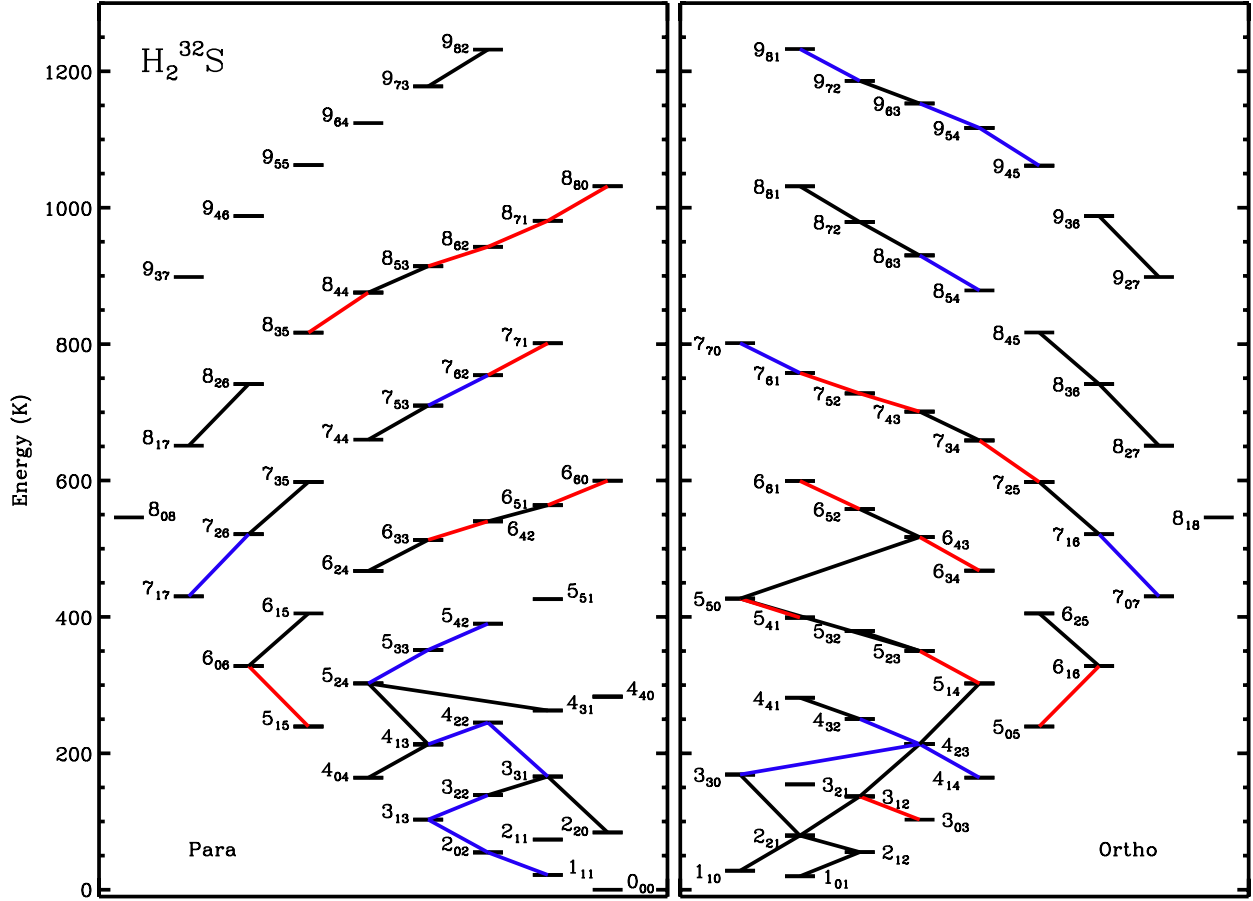}
\caption{The right and left panels plot energy level diagrams for ortho and para \htsi, respectively. Transitions detected in the HIFI scan are indicated by lines connecting upper and lower states and are color coded according the how blended each transition is. Black, blue, and red correspond to categories 1 (not blended), 2 (partially blended), and 3 (heavily blended), respectively. See Sec.~\ref{s-mli} for more details on how these categories are defined. \label{p-gd}}
\end{figure}

\clearpage

\begin{figure}
\plotone{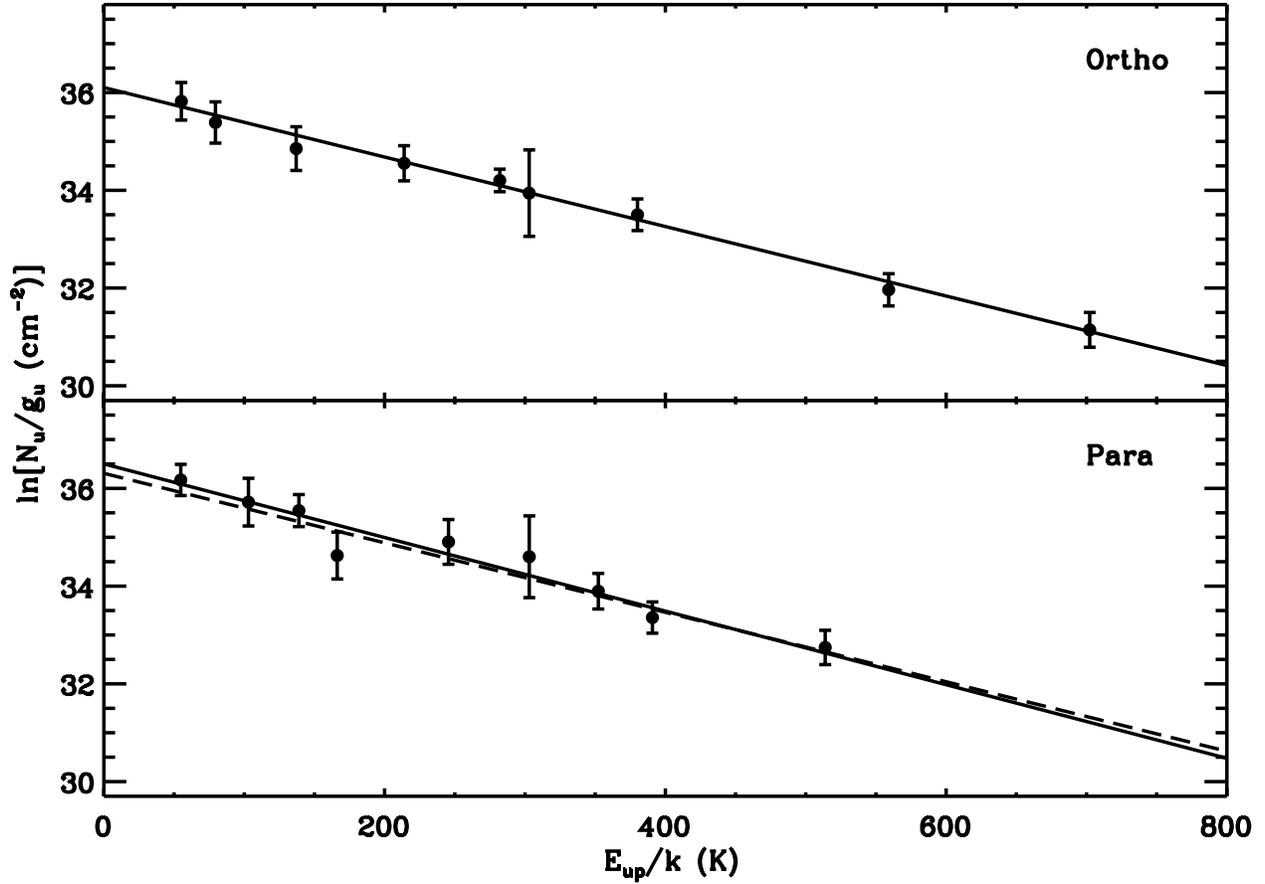}
\caption{Rotation diagrams for ortho and para states where \nup\ values could be computed are given in the upper and lower panels, respectively. All points were derived assuming a source size of 6\arcsec\ and Orion KL isotopic ratios. Independent linear least squares fits to the ortho and para points are plotted as solid lines in either panel. The dashed line in the bottom panel is a fit to the para points with \trot\ fixed at 141~K, the rotation temperature derived from ortho \hts. \label{p-rd}}
\end{figure}

\clearpage

\begin{figure}
\epsscale{1.0}
\plotone{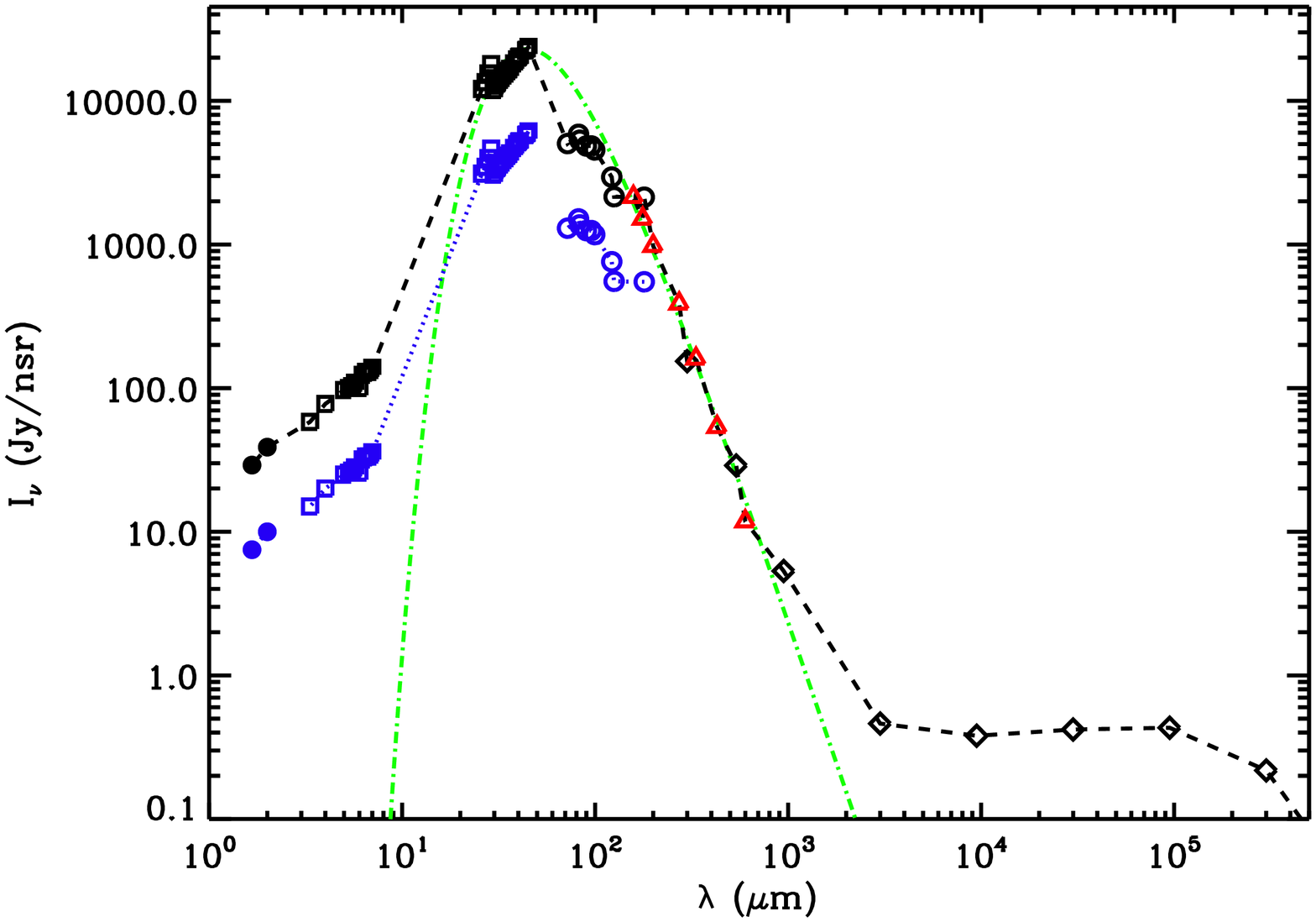}
\caption{The observed continuum toward IRc2. The red triangles indicate the continuum level as measured in the HIFI scan. The blue squares and circles (not filled in) correspond to flux measurements obtained from ISO SWS \citep{vandishoeck98} and LWS \citep{lerate06}, respectively. The blue dots (filled in) represent observations from \citet{lonsdale82}. The black squares, circles, and dots are the same measurements as their respective blue counterparts normalized so that the LWS observations agree with the HIFI continuum at $\sim$158~\um. Black diamonds correspond to flux measurements presented by \citet{dicker09}. The green dot-dashed line is a 65~K blackbody. The black dashed line represents the ``observed radiation field" used in the RADEX models described in the text. \label{p-bg_rad}}
\end{figure}

\clearpage

\begin{figure}
\includegraphics[angle=90, scale=0.70]{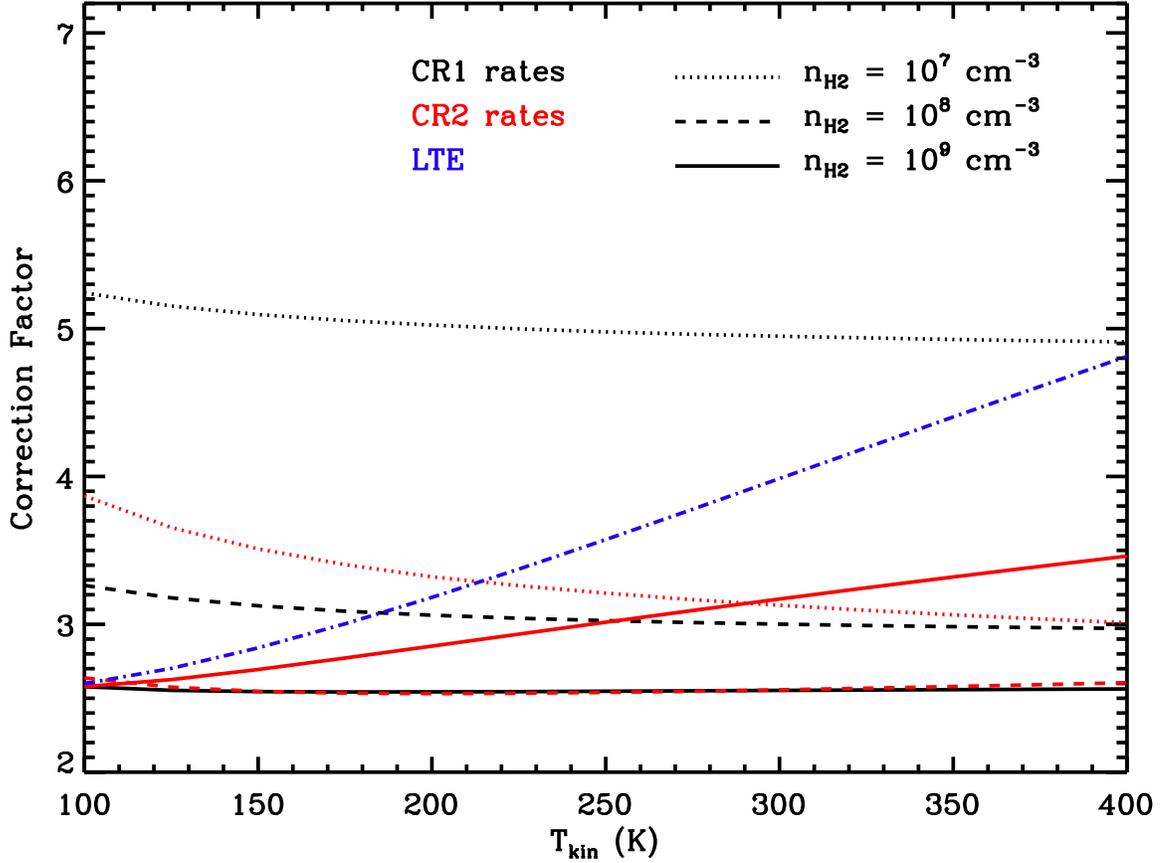}
\caption{Correction factors which convert \nobs\ to \ntot\ plotted as a function of kinetic temperature. Correction factors that assume LTE are plotted as a blue, dot-dashed line. CF values computed using RADEX assuming the CR1 (black) and CR2 (red) collision rates are also plotted with different densities represented as different types of lines. The RADEX models set \ntot~=~5.0\sn$^{16}$~\cms, \dv~=~8.6~km/s, and assume the enhanced continuum. \label{cf}}
\end{figure}

\clearpage

\begin{figure}
\epsscale{0.8}
\plotone{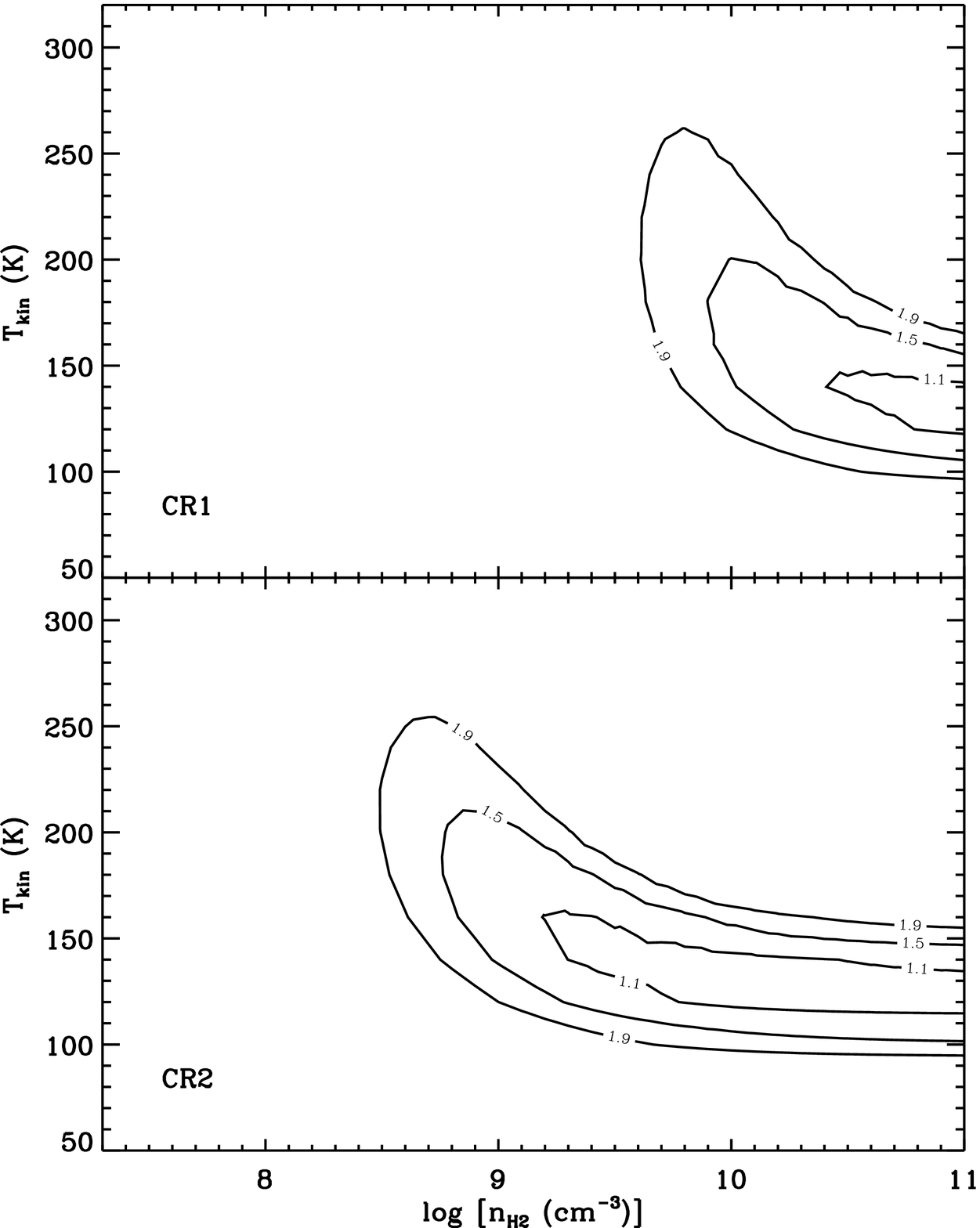}
\caption{The upper and lower panels are \redchisq\ contour plots comparing \htsii\ and \htsiii\ emission to RADEX model grids assuming CR1 and CR2 collision rates, respectively. All model realizations assume the observed radiation field, and set \ssize~=~6\arcsec and \dv~=~8.6~km/s. We also set \ntot(\htsi)~=~9.5\sn$^{17}$~\cms, but scale the isotopic emission according to Orion KL ratios. The \redchisq\ values corresponding to each contour are labeled in the plot. \label{p-chisq_orion}}
\end{figure}

\clearpage

\begin{figure}
\epsscale{0.61}
\plotone{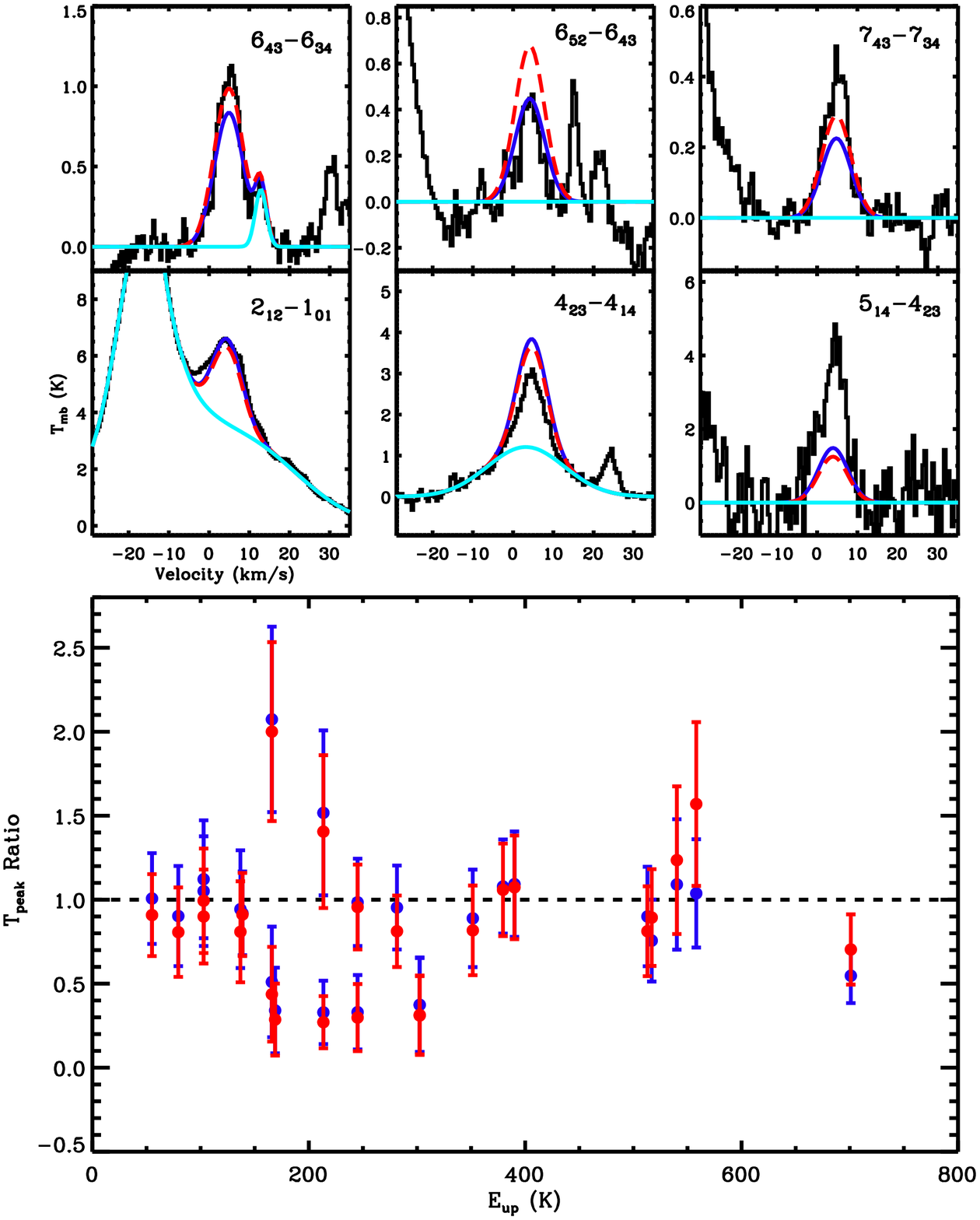}
\caption{Comparison of the observed \htsiii\ lines with RADEX models. The large bottom panel plots the ratio of predicted to observed \tpeak\ as a function of \eup. Blue and red points represent \tpeak\ ratios computed using RADEX models that fit the data well (\redchisq~$<$~1.1) assuming the observed (\nht~=~7.0\sn$^{9}$~\cmc, \tkin~=~130~K) and enhanced (\nht~=~3.0\sn$^{9}$~\cmc, \tkin~=~120~K) radiation fields, respectively. The upper panels plot a sample of \htsiii\ lines in black with the same models overlaid using the same color conventions as the lower panel (the red line is dashed because it overlaps with the blue in some instances).  Blending lines and spatial/velocity components other than the hot core, fit with Gaussians, are overlaid in cyan. Model line profiles are added to the components represented in cyan. Transitions increase in excitation from the lower left panel to the upper right. The models use CR2 collision rates and set \ssize~=~6\arcsec, \dv~=~8.6~km/s, and \ntot(\htsiii)~=~4.8\sn$^{16}$~\cms. \label{p-he34}}
\end{figure}

\clearpage

\begin{figure}
\epsscale{0.67}
\plotone{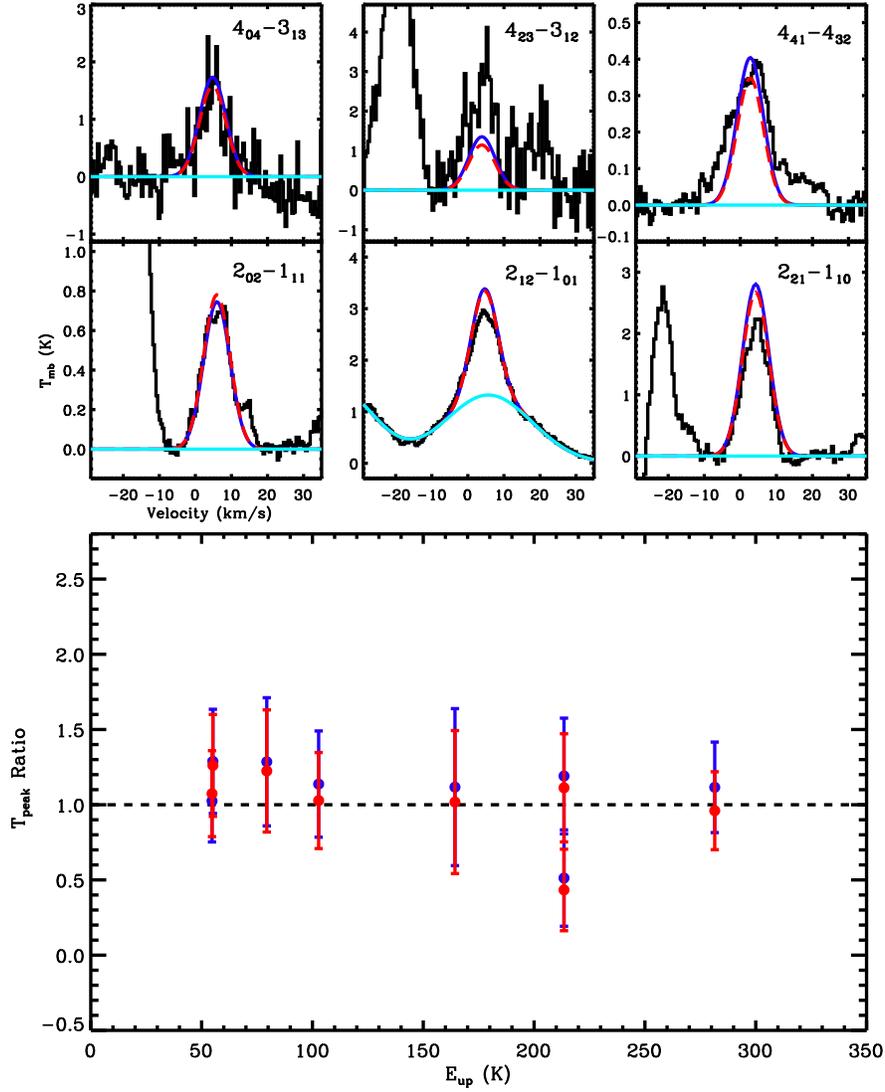}
\caption{Same as Fig.~\ref{p-he34} for \htsii\ lines. The same RADEX model parameters used in Fig.~\ref{p-he34} are used here for both the observed (blue) and enhanced (red) radiation fields except for the total column density which is set to \ntot(\htsii)~=~1.3\sn$^{16}$~\cms, in agreement with Orion~KL isotopic ratios. \label{p-he33}}
\end{figure}

\clearpage

\begin{figure}
\epsscale{0.75}
\plotone{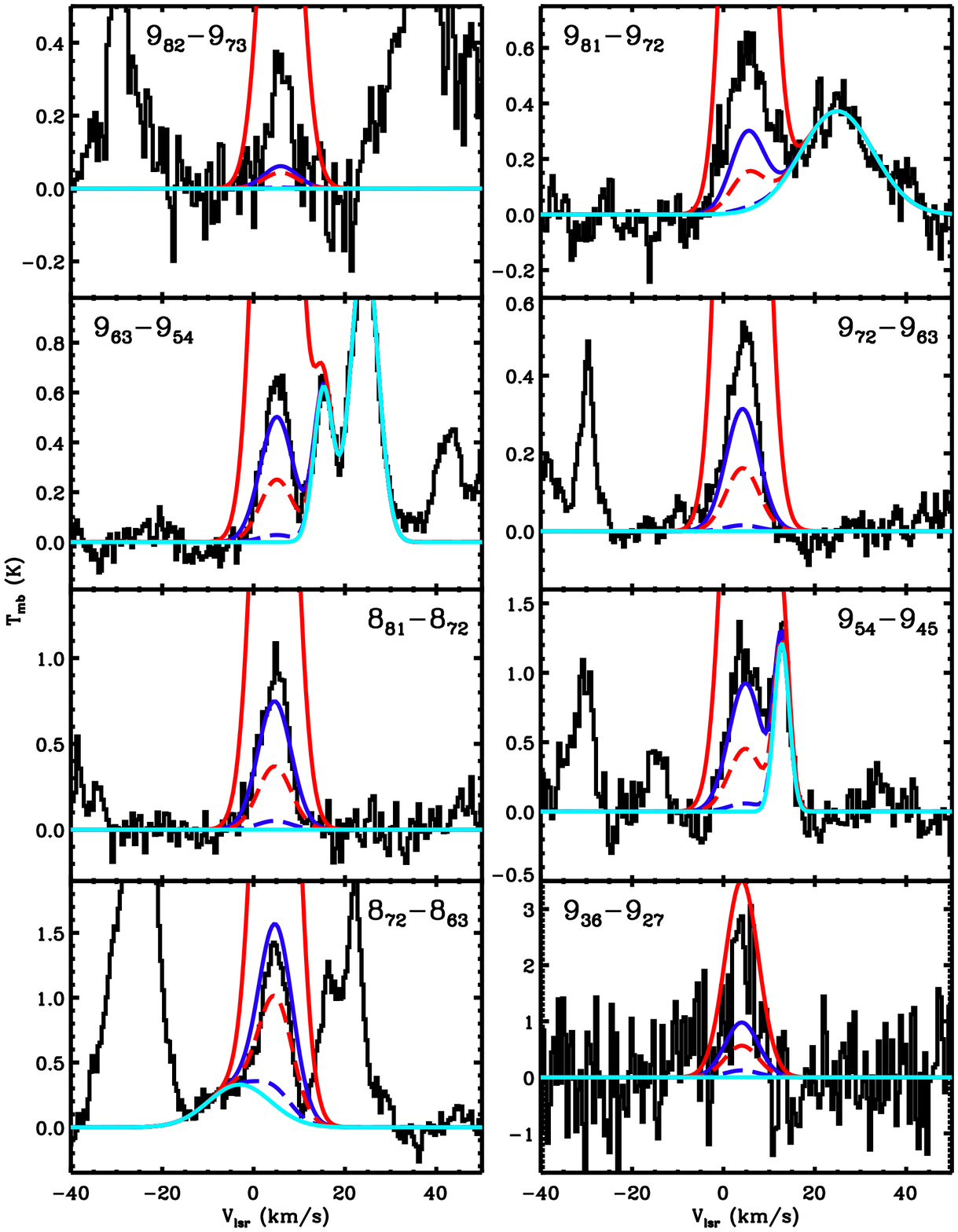}
\caption{Each panel plots a highly excited \htsi\ line in black. The transition is given in the upper part of each panel. Blending lines fit with Gaussians are overlaid in cyan. RADEX models are overlaid as different colors and model line profiles are added to the components represented in cyan. Blue and red lines correspond to models with \nht~=~1.5\sn$^{10}$ and 1.5\sn$^{11}$~\cmc, while dashed and solid lines correspond to \tkin~=~140 and 300~K, respectively. All models set \ssize~=~6\arcsec, \dv~=~8.6~km/s, \ntot(\htsi)~=~9.5\sn$^{17}$~\cms, use CR1 collision rates, and assume the observed radiation field. \label{p-he_orion}}
\end{figure}

\clearpage

\begin{figure}
\epsscale{0.8}
\plotone{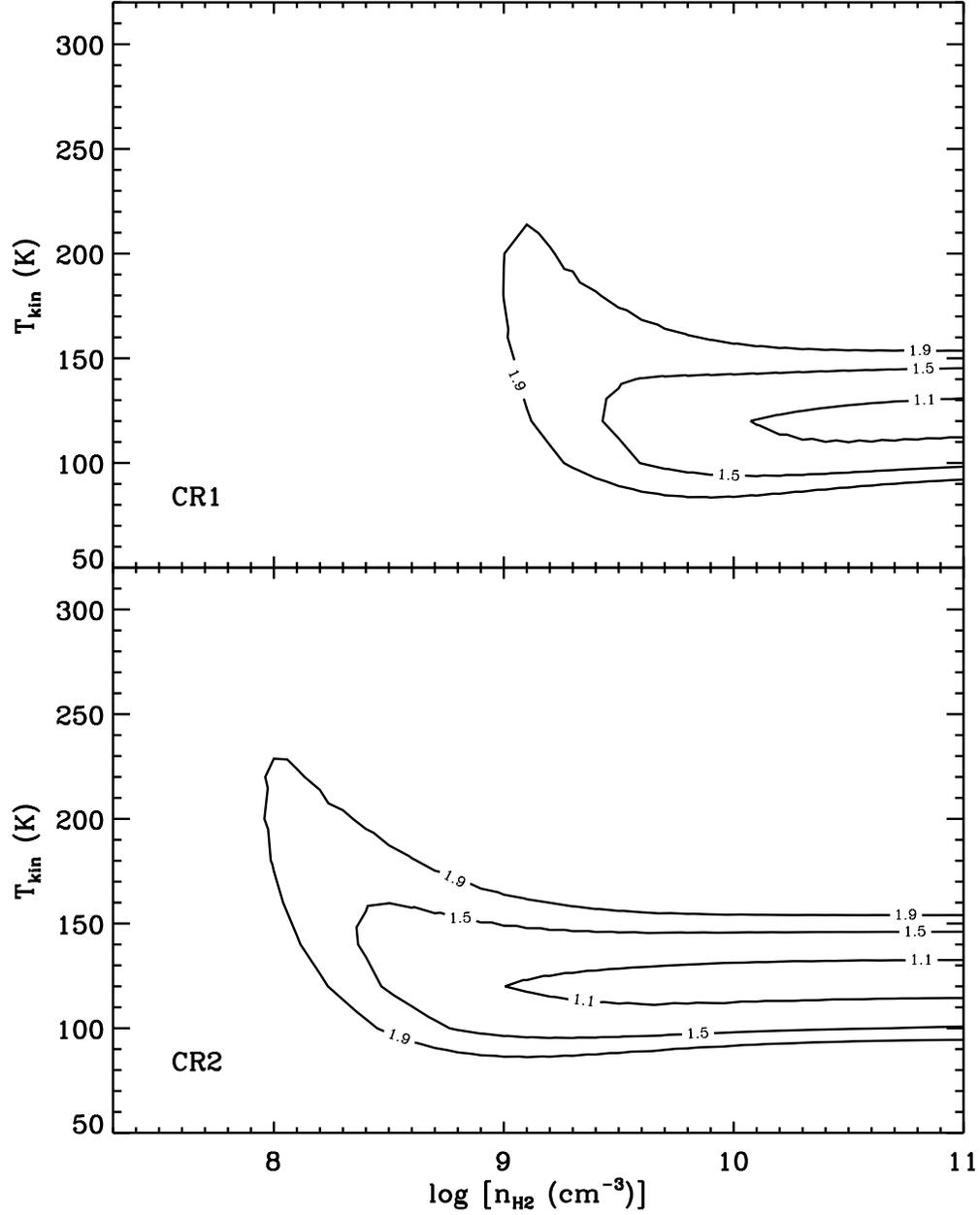}
\caption{The upper and lower panels are \redchisq\ contour plots computed using the same methodology as Fig.~\ref{p-chisq_orion}, except the RADEX model grids used here assume a radiation field that is enhanced by a factor of 8 for $\lambda$~$<$~100~\um. The \redchisq\ values corresponding to each contour are labeled in the plot. \label{p-chisq_ir100sc30}}
\end{figure}

\clearpage

\begin{figure}
\epsscale{0.8}
\plotone{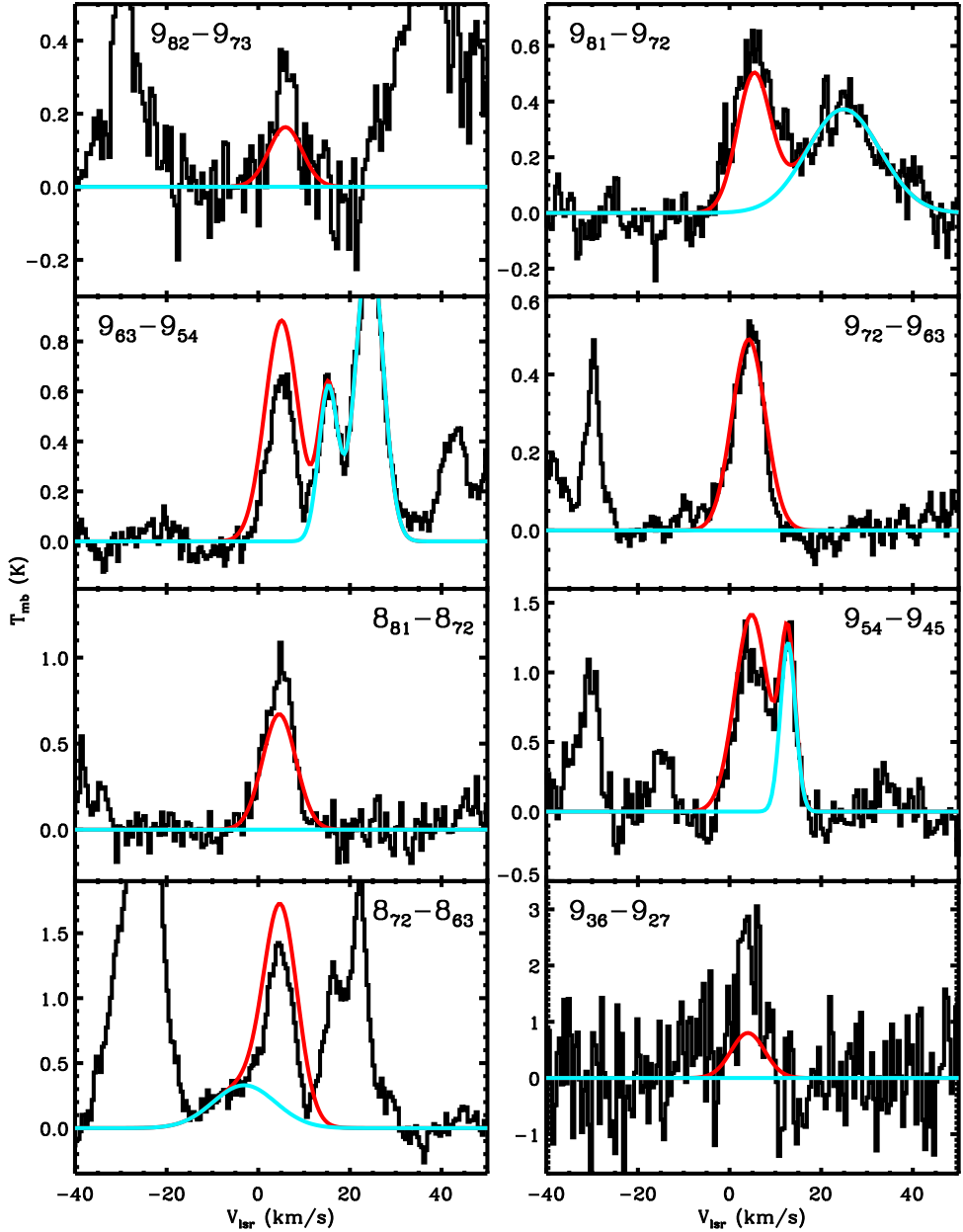}
\caption{The same \htsi\ transitions as Fig.~\ref{p-he_orion} with one RADEX model overlaid in red. Blending lines fit with Gaussians are overlaid in cyan. Model line profiles are added to the components represented in cyan. The RADEX model sets \ssize~=~6\arcsec, \dv~=~8.6~km/s, \nht~=~3.0\sn$^{9}$~\cmc, \tkin~=~120~K, and assumes the enhanced continuum. \label{p-he_sc}}
\end{figure}

\clearpage

\begin{figure}
\epsscale{0.8}
\plotone{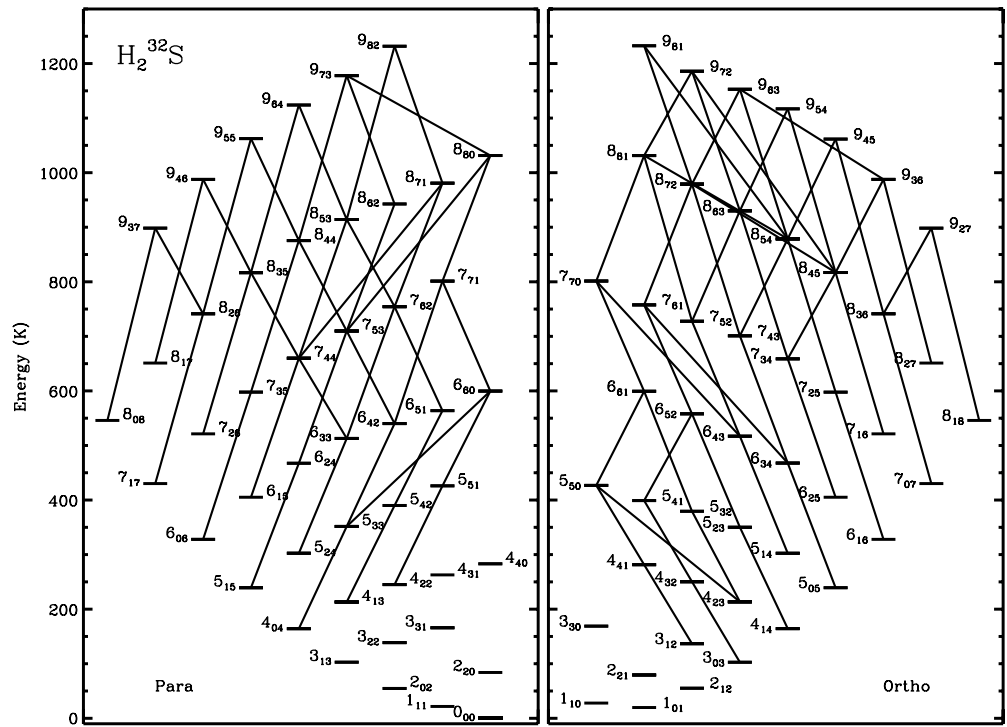}
\caption{The same energy level diagram as Fig.~\ref{p-gd} is plotted, except lines connecting the states represent transitions with $\lambda$~$<$~100~\um\ and $\mu^{2}$S~$>$~0.01. \label{p-gd_irpump}}
\end{figure}

\clearpage

\begin{figure}
\epsscale{0.8}
\plotone{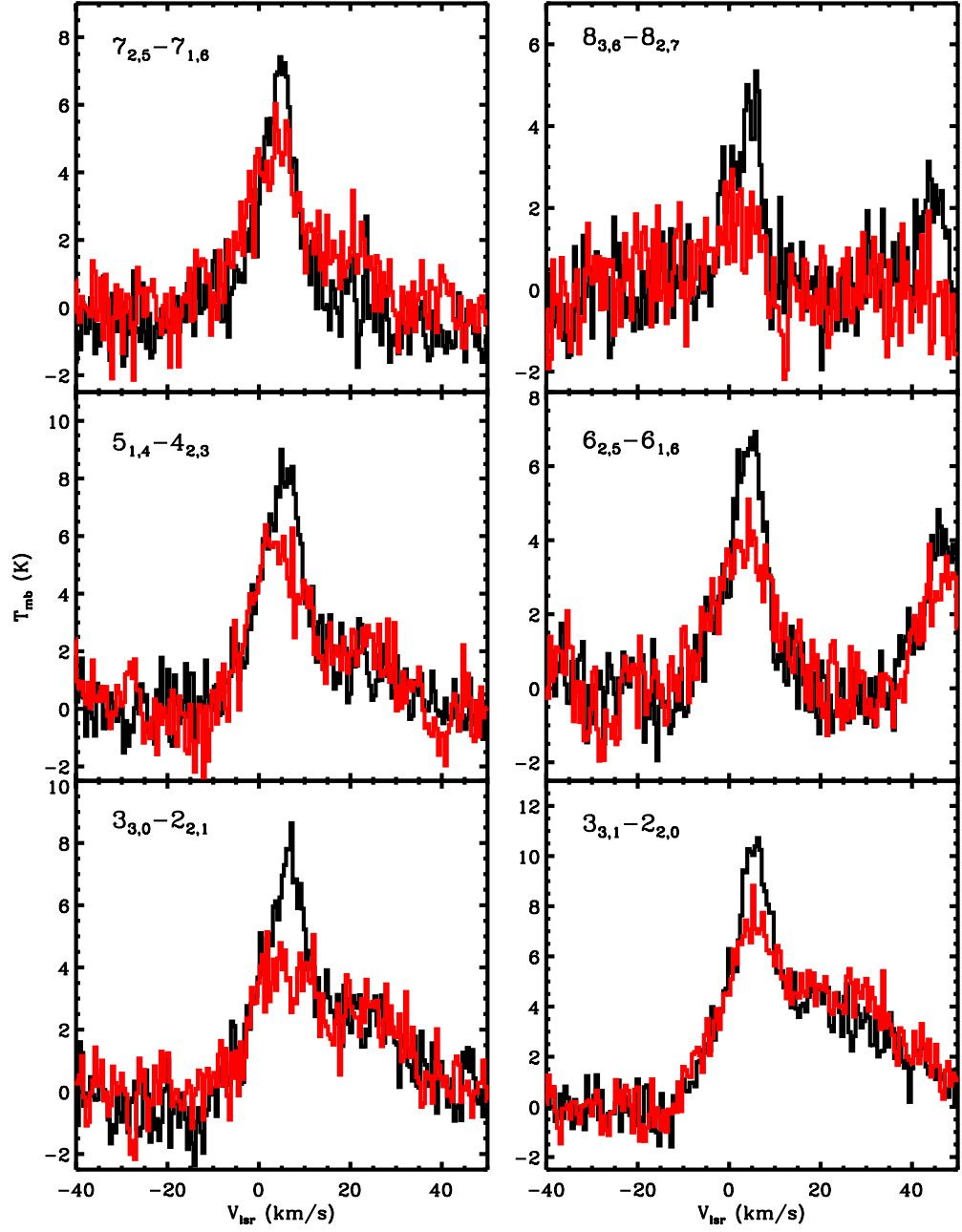}
\caption{\htsi\ lines from bands 6 and 7 are plotted for the hot core (black) and compact ridge (red) pointings. The lines are clearly stronger toward the hot core. \label{p-tpoint}}
\end{figure}

\clearpage

\begin{deluxetable}{lccccccc}
\tablecaption{Measured line parameters for the hot core \label{t-hc}}
\tablecolumns{8}
\tablewidth{0pt}
\tablehead{\colhead{Transition} & \colhead{$\nu$} & \colhead{E$_{up}$} & \colhead{v$_{LSR}$} & \colhead{$\Delta$v}  &
\colhead{T$_{peak}$} & \colhead{\tint} & \colhead{Notes\tablenotemark{a}} \\
\colhead{(J$_{K^{-},K^{+}}$)} & \colhead{(GHz)} & \colhead{(K)} & \colhead{(km s$^{-1}$)}   & \colhead{(km s$^{-1}$)} & \colhead{(K)} & \colhead{(K km s$^{-1}$)}  & \colhead{}}
\startdata
\cutinhead{\htsi}
2$_{0,2}$--1$_{1,1}$     &      687.30  &        54.7  &      4.5   $\pm$     0.2  &     10.3   $\pm$     0.4  &     3.28   $\pm$    0.03  &      35.8   $\pm$     2.4  &   2 \\     
2$_{1,2}$--1$_{0,1}$     &      736.03  &        55.1  &      4.9   $\pm$     0.1  &      8.9   $\pm$     0.1  &     2.32   $\pm$    0.04  &      21.8   $\pm$     0.3  &   1 \\     
2$_{2,1}$--2$_{1,2}$     &      505.57  &        79.4  &      4.4   $\pm$     0.6  &     11.4   $\pm$     0.6  &     2.01   $\pm$    0.02  &      24.3   $\pm$     1.6  &   1 \\     
2$_{2,1}$--1$_{1,0}$     &     1072.84  &        79.4  &      3.9   $\pm$     0.1  &      8.9   $\pm$     0.1  &     3.42   $\pm$    0.07  &      32.4   $\pm$     0.2  &   1 \\     
3$_{1,3}$--2$_{0,2}$     &     1002.78  &       102.9  &      5.4   $\pm$     0.1  &     13.2   $\pm$     0.2  &     3.69   $\pm$    0.1  &      51.9   $\pm$     1.2  &   2 \\     
3$_{1,2}$--2$_{2,1}$     &     1196.01  &       136.8  &      4.5   $\pm$     0.1  &     10.1   $\pm$     0.2  &      5.9   $\pm$     0.2  &      63.4   $\pm$     1.7  &   1 \\     
3$_{2,2}$--3$_{1,3}$     &      747.30  &       138.7  &      4.9   $\pm$     0.1  &      9.9   $\pm$     0.2  &     3.57   $\pm$    0.05  &      37.5   $\pm$     0.9  &   2 \\     
3$_{3,1}$--3$_{2,2}$     &      568.05  &       166.0  &      4.6   $\pm$     0.1  &      9.0   $\pm$     0.1  &     1.58   $\pm$    0.02  &      15.2   $\pm$     0.2  &   1 \\     
3$_{3,1}$--2$_{2,0}$     &     1707.97  &       166.0  &      5.4   $\pm$     0.1  &      6.7   $\pm$     0.4  &      5.9   $\pm$     0.6  &      42.0   $\pm$     2.9  &   1 \\     
3$_{3,0}$--2$_{2,1}$     &     1865.62  &       169.0  &      5.9   $\pm$     0.3  &      7.6   $\pm$     0.8  &      4.2   $\pm$     0.7  &      33.7   $\pm$     4.2  &   1 \\     
4$_{1,3}$--4$_{0,4}$     &     1018.35  &       213.2  &      4.2   $\pm$     0.1  &      9.3   $\pm$     0.3  &     4.05   $\pm$    0.09  &      40.0   $\pm$     2.2  &   1 \\     
4$_{2,3}$--3$_{3,0}$     &      930.15  &       213.6  &      3.9   $\pm$     0.1  &      8.7   $\pm$     0.1  &     2.78   $\pm$    0.06  &      25.9   $\pm$     0.1  &   2 \\     
4$_{2,3}$--4$_{1,4}$     &     1026.51  &       213.6  &      3.8   $\pm$     0.4  &      8.3   $\pm$     0.4  &      4.3   $\pm$     0.1  &      38.2   $\pm$     2.0  &   2 \\     
4$_{2,3}$--3$_{1,2}$     &     1599.75  &       213.6  &      4.5   $\pm$     0.2  &      9.7   $\pm$     0.5  &      6.6   $\pm$     0.6  &      67.9   $\pm$     5.0  &   1 \\     
4$_{2,2}$--4$_{1,3}$     &      665.39  &       245.1  &      4.1   $\pm$     0.5  &     10.4   $\pm$     0.5  &     2.75   $\pm$    0.05  &      30.4   $\pm$     1.4  &   2 \\     
4$_{2,2}$--3$_{3,1}$     &     1648.71  &       245.2  &      3.8   $\pm$     0.2  &      7.1   $\pm$     0.6  &      6.5   $\pm$     0.9  &      49.4   $\pm$     6.6  &   2 \\     
4$_{3,2}$--4$_{2,3}$     &      765.94  &       250.3  &      3.2   $\pm$     0.4  &      9.7   $\pm$     0.3  &     3.16   $\pm$    0.03  &      32.7   $\pm$     2.8  &   2 \\     
4$_{4,1}$--4$_{3,2}$     &      650.37  &       281.5  &      4.1   $\pm$     0.1  &      9.4   $\pm$     0.1  &     2.38   $\pm$    0.03  &      23.7   $\pm$     0.3  &   1 \\     
5$_{1,4}$--4$_{2,3}$     &     1852.69  &       302.5  &      5.0   $\pm$     0.2  &      9.3   $\pm$     0.6  &      6.3   $\pm$     0.7  &      62.4   $\pm$     5.0  &   1 \\     
5$_{2,4}$--4$_{3,1}$     &      827.92  &       302.6  &      4.2   $\pm$     0.3  &      7.1   $\pm$     0.8  &     0.37   $\pm$    0.06  &       2.8   $\pm$     0.2  &   1 \\     
5$_{2,4}$--4$_{1,3}$     &     1862.44  &       302.6  &      5.1   $\pm$     0.2  &      9.5   $\pm$     0.5  &      6.5   $\pm$     0.8  &      66.0   $\pm$     2.6  &   1 \\     
6$_{0,6}$--5$_{1,5}$     &     1846.74  &       328.1  &      2.8   $\pm$     0.2  &     11.2   $\pm$     0.6  &      7.9   $\pm$     0.7  &      94.1   $\pm$     7.9  &   3 \\     
5$_{3,3}$--5$_{2,4}$     &     1023.12  &       351.7  &      3.9   $\pm$     0.1  &     10.4   $\pm$     0.1  &      4.5   $\pm$     0.1  &      49.9   $\pm$     0.7  &   2 \\     
5$_{3,2}$--5$_{2,3}$     &      611.44  &       379.5  &      4.1   $\pm$     0.1  &     10.7   $\pm$     0.4  &     2.37   $\pm$    0.02  &      27.0   $\pm$     1.6  &   1 \\     
5$_{4,2}$--5$_{3,3}$     &      800.85  &       390.1  &      4.0   $\pm$     0.6  &     10.2   $\pm$     0.6  &     2.98   $\pm$    0.09  &      32.3   $\pm$     1.3  &   2 \\     
6$_{1,5}$--6$_{0,6}$     &     1608.37  &       405.2  &      3.9   $\pm$     0.3  &      9.7   $\pm$     0.6  &      4.3   $\pm$     0.8  &      44.1   $\pm$     2.4  &   1 \\     
6$_{2,5}$--6$_{1,6}$     &     1608.60  &       405.2  &      3.9   $\pm$     0.2  &      9.7   $\pm$     0.5  &      6.4   $\pm$     0.8  &      65.7   $\pm$     2.7  &   1 \\     
5$_{5,0}$--5$_{2,3}$     &     1598.92  &       426.9  &      4.2   $\pm$     0.8  &     15.5   $\pm$     2.5  &      2.0   $\pm$     0.7  &      33.4   $\pm$     4.1  &   1 \\     
6$_{3,3}$--6$_{2,4}$     &      947.26  &       513.0  &      3.5   $\pm$     0.1  &     11.2   $\pm$     0.3  &      2.8   $\pm$     0.2  &      32.9   $\pm$     1.1  &   1 \\     
6$_{4,3}$--5$_{5,0}$     &     1879.36  &       517.1  &      4.3   $\pm$     0.4  &      4.0   $\pm$     0.8  &      2.2   $\pm$     0.9  &       9.4   $\pm$     1.7  &   1 \\     
7$_{1,6}$--7$_{0,7}$     &     1900.14  &       521.4  &      5.2   $\pm$     0.8  &      6.0   $\pm$     1.0  &      5.4   $\pm$     0.9  &      34.3   $\pm$    11.8  &   2 \\     
7$_{2,6}$--7$_{1,7}$     &     1900.18  &       521.4  &      4.8   $\pm$     1.0  &      6.9   $\pm$     1.9  &      5.1   $\pm$     0.9  &      37.7   $\pm$    12.8  &   2 \\     
6$_{5,2}$--6$_{4,3}$     &      854.97  &       558.1  &      3.6   $\pm$     0.1  &     10.6   $\pm$     0.1  &     3.36   $\pm$    0.05  &      37.8   $\pm$     0.5  &   1 \\     
6$_{5,1}$--6$_{4,2}$     &      493.36  &       563.9  &      4.8   $\pm$     0.1  &      6.1   $\pm$     0.2  &     0.52   $\pm$    0.01  &       3.4   $\pm$     0.2  &   1 \\     
7$_{2,5}$--7$_{1,6}$     &     1592.67  &       597.9  &      4.3   $\pm$     0.2  &      9.2   $\pm$     0.5  &      6.6   $\pm$     0.9  &      64.4   $\pm$     2.8  &   1 \\     
7$_{3,5}$--7$_{2,6}$     &     1593.97  &       597.9  &      2.3   $\pm$     0.3  &     11.9   $\pm$     0.8  &      4.5   $\pm$     0.8  &      57.2   $\pm$     3.1  &   1 \\     
7$_{4,3}$--7$_{3,4}$     &      880.06  &       701.1  &      4.4   $\pm$     0.1  &      6.8   $\pm$     0.4  &     1.77   $\pm$    0.05  &      12.9   $\pm$     1.3  &   1 \\     
7$_{5,3}$--7$_{4,4}$     &     1040.28  &       709.9  &      4.2   $\pm$     0.1  &      8.1   $\pm$     0.3  &      2.0   $\pm$     0.2  &      17.6   $\pm$     0.5  &   1 \\     
8$_{2,6}$--8$_{1,7}$     &     1882.52  &       741.5  &      4.6   $\pm$     0.3  &      5.7   $\pm$     0.9  &      2.4   $\pm$     0.7  &      14.8   $\pm$     1.8  &   1 \\     
8$_{3,6}$--8$_{2,7}$     &     1882.77  &       741.5  &      3.6   $\pm$     0.3  &      8.7   $\pm$     0.6  &      4.2   $\pm$     0.8  &      38.6   $\pm$     2.5  &   1 \\     
7$_{6,2}$--7$_{5,3}$     &      928.64  &       754.5  &      4.3   $\pm$     0.1  &      8.2   $\pm$     0.1  &     1.91   $\pm$    0.06  &      16.7   $\pm$     0.2  &   2 \\     
7$_{7,0}$--7$_{6,1}$     &      910.67  &       801.5  &      4.7   $\pm$     0.1  &      7.1   $\pm$     0.2  &     1.45   $\pm$    0.06  &      10.9   $\pm$     0.4  &   2 \\     
8$_{4,5}$--8$_{3,6}$     &     1576.44  &       817.1  &      3.3   $\pm$     0.3  &      7.2   $\pm$     0.7  &      4.2   $\pm$     0.9  &      32.4   $\pm$     2.6  &   1 \\     
8$_{5,3}$--8$_{4,4}$     &      804.73  &       914.2  &      4.7   $\pm$     0.1  &      8.5   $\pm$     0.4  &     0.76   $\pm$    0.07  &       6.9   $\pm$     0.2  &   1 \\     
8$_{6,3}$--8$_{5,4}$     &     1071.31  &       930.2  &      4.5   $\pm$     0.1  &      8.5   $\pm$     0.2  &      2.4   $\pm$     0.1  &      22.2   $\pm$     0.4  &   2 \\     
8$_{7,2}$--8$_{6,3}$     &     1019.45  &       979.1  &      4.9   $\pm$     0.1  &      6.5   $\pm$     0.2  &     1.23   $\pm$    0.09  &       8.5   $\pm$     0.6  &   1 \\     
9$_{3,6}$--9$_{2,7}$     &     1860.98  &       987.8  &      4.0   $\pm$     0.4  &      6.4   $\pm$     0.8  &      2.5   $\pm$     0.7  &      17.0   $\pm$     2.0  &   1 \\     
8$_{8,1}$--8$_{7,2}$     &     1091.26  &      1031.5  &      4.6   $\pm$     0.1  &      7.1   $\pm$     0.3  &     0.93   $\pm$    0.08  &       7.1   $\pm$     0.2  &   1 \\     
9$_{5,4}$--9$_{4,5}$     &     1154.68  &      1117.1  &      4.8   $\pm$     0.3  &      8.1   $\pm$     0.8  &      1.2   $\pm$     0.1  &      10.0   $\pm$     0.8  &   2 \\     
9$_{6,3}$--9$_{5,4}$     &      746.73  &      1152.9  &      5.1   $\pm$     0.1  &      6.7   $\pm$     0.2  &     0.65   $\pm$    0.05  &       4.7   $\pm$     0.1  &   2 \\     
9$_{7,2}$--9$_{6,3}$     &      689.12  &      1186.0  &      4.2   $\pm$     0.1  &      7.5   $\pm$     0.2  &     0.51   $\pm$    0.03  &       4.0   $\pm$     0.1  &   1 \\     
9$_{8,2}$--9$_{7,3}$     &     1122.64  &      1231.9  &      5.9   $\pm$     0.3  &      5.4   $\pm$     0.8  &     0.34   $\pm$    0.09  &       1.9   $\pm$     0.2  &   1 \\     
9$_{8,1}$--9$_{7,2}$     &      973.85  &      1232.7  &      5.3   $\pm$     0.2  &     10.1   $\pm$     0.5  &     0.56   $\pm$    0.07  &       6.0   $\pm$     0.3  &   2 \\     
\cutinhead{\htsiii}
2$_{1,2}$--1$_{0,1}$     &      734.27  &        55.0  &      4.5   $\pm$     0.2  &     10.2   $\pm$     0.4  &     2.96   $\pm$    0.04  &      32.0   $\pm$     2.5  &   2 \\     
2$_{2,1}$--1$_{1,0}$     &     1069.37  &        79.2  &      3.7   $\pm$     0.1  &     10.3   $\pm$     0.1  &     4.00   $\pm$    0.07  &      44.0   $\pm$     0.9  &   2 \\     
3$_{0,3}$--2$_{1,2}$     &      991.73  &       102.7  &      4.0   $\pm$     0.1  &      9.2   $\pm$     0.2  &     3.49   $\pm$    0.08  &      34.1   $\pm$     1.2  &   2 \\     
3$_{1,3}$--2$_{0,2}$     &     1000.91  &       102.7  &      3.8   $\pm$     0.1  &      7.6   $\pm$     0.1  &     2.95   $\pm$    0.08  &      23.8   $\pm$     0.5  &   2 \\     
3$_{1,2}$--2$_{2,1}$     &     1197.18  &       136.7  &      3.7   $\pm$     0.1  &     11.1   $\pm$     0.5  &      3.4   $\pm$     0.1  &      40.7   $\pm$     3.1  &   2 \\     
3$_{2,2}$--3$_{1,3}$     &      745.52  &       138.5  &      3.9   $\pm$     0.1  &      9.7   $\pm$     0.1  &     1.30   $\pm$    0.04  &      13.4   $\pm$     0.1  &   1 \\     
3$_{3,1}$--3$_{2,2}$     &      563.68  &       165.5  &      4.7   $\pm$     0.2  &      5.2   $\pm$     0.5  &     0.33   $\pm$    0.02  &       1.9   $\pm$     0.3  &   1 \\     
3$_{3,1}$--2$_{2,0}$     &     1702.01  &       165.6  &      4.0   $\pm$     0.3  &      7.9   $\pm$     0.6  &      3.5   $\pm$     0.6  &      29.2   $\pm$     2.0  &   1 \\     
3$_{3,0}$--2$_{2,1}$     &     1861.85  &       168.6  &      5.0   $\pm$     0.2  &      7.2   $\pm$     0.6  &      4.3   $\pm$     0.8  &      32.5   $\pm$     2.2  &   1 \\     
4$_{2,3}$--4$_{1,4}$     &     1024.85  &       213.2  &      4.7   $\pm$     0.1  &      7.3   $\pm$     0.3  &      1.7   $\pm$     0.1  &      13.5   $\pm$     0.9  &   2 \\     
4$_{2,3}$--3$_{1,2}$     &     1595.98  &       213.3  &      4.8   $\pm$     0.2  &      9.4   $\pm$     0.6  &      5.4   $\pm$     0.8  &      54.3   $\pm$     2.8  &   1 \\     
4$_{2,2}$--4$_{1,3}$     &      666.82  &       244.9  &      4.5   $\pm$     0.1  &      8.4   $\pm$     0.3  &     0.87   $\pm$    0.03  &       7.8   $\pm$     0.3  &   2 \\     
4$_{2,2}$--3$_{3,1}$     &     1653.14  &       244.9  &      3.3   $\pm$     0.7  &     10.8   $\pm$     2.0  &      1.9   $\pm$     0.6  &      21.5   $\pm$     2.8  &   1 \\     
4$_{4,1}$--4$_{3,2}$     &      643.59  &       280.7  &      4.7   $\pm$     0.1  &      8.0   $\pm$     0.2  &     1.21   $\pm$    0.03  &      10.3   $\pm$     0.4  &   2 \\     
5$_{1,4}$--4$_{2,3}$     &     1849.96  &       302.1  &      4.0   $\pm$     0.3  &      7.7   $\pm$     0.7  &      4.0   $\pm$     0.8  &      32.4   $\pm$     2.4  &   1 \\     
5$_{2,4}$--4$_{1,3}$     &     1859.05  &       302.1  &      5.6   $\pm$     0.2  &      4.7   $\pm$     0.7  &      3.0   $\pm$     0.6  &      15.0   $\pm$     1.6  &   1 \\     
6$_{0,6}$--5$_{1,5}$     &     1843.77  &       327.5  &      3.8   $\pm$     0.2  &      7.9   $\pm$     0.4  &      4.4   $\pm$     0.8  &      37.4   $\pm$     0.5  &   3 \\     
5$_{3,3}$--5$_{2,4}$     &     1021.31  &       351.1  &      4.5   $\pm$     0.5  &      7.2   $\pm$     0.7  &     0.90   $\pm$    0.07  &       6.9   $\pm$     1.8  &   2 \\     
5$_{3,2}$--5$_{2,3}$     &      613.83  &       379.1  &      4.5   $\pm$     0.1  &      8.3   $\pm$     0.1  &     0.91   $\pm$    0.02  &       8.0   $\pm$     0.1  &   2 \\     
5$_{4,2}$--5$_{3,3}$     &      796.54  &       389.3  &      4.0   $\pm$     0.1  &      6.2   $\pm$     0.3  &     0.44   $\pm$    0.04  &       2.9   $\pm$     0.1  &   2 \\     
6$_{3,3}$--6$_{2,4}$     &      949.42  &       512.4  &      4.9   $\pm$     0.2  &      6.2   $\pm$     0.5  &     0.35   $\pm$    0.05  &       2.3   $\pm$     0.2  &   1 \\     
6$_{4,3}$--6$_{3,4}$     &     1023.49  &       516.3  &      4.9   $\pm$     0.1  &      6.8   $\pm$     0.2  &     1.10   $\pm$    0.06  &       7.9   $\pm$     0.2  &   1 \\     
6$_{4,2}$--6$_{3,3}$     &      568.82  &       539.7  &      3.4   $\pm$     0.3  &      5.5   $\pm$     0.6  &     0.12   $\pm$    0.03  &       0.7   $\pm$     0.1  &   2 \\     
6$_{5,2}$--6$_{4,3}$     &      848.36  &       557.0  &      4.1   $\pm$     0.2  &      5.9   $\pm$     0.4  &     0.43   $\pm$    0.06  &       2.7   $\pm$     0.2  &   1 \\     
7$_{4,3}$--7$_{3,4}$     &      884.52  &       700.4  &      4.8   $\pm$     0.1  &      6.8   $\pm$     0.2  &     0.41   $\pm$    0.03  &       3.0   $\pm$     0.1  &   1 \\     
\cutinhead{\htsii}
2$_{0,2}$--1$_{1,1}$     &      687.16  &        54.7  &      6.0   $\pm$     0.1  &      9.2   $\pm$     0.1  &     0.73   $\pm$    0.02  &       7.1   $\pm$     0.1  &   2 \\     
2$_{1,2}$--1$_{0,1}$     &      735.13  &        55.1  &      4.7   $\pm$     0.1  &      9.5   $\pm$     0.1  &     1.60   $\pm$    0.02  &      16.2   $\pm$     0.3  &   2 \\     
2$_{2,1}$--1$_{1,0}$     &     1071.05  &        79.3  &      4.3   $\pm$     0.1  &      8.2   $\pm$     0.1  &     2.18   $\pm$    0.07  &      19.1   $\pm$     0.2  &   2 \\     
3$_{0,3}$--2$_{1,2}$     &      992.40  &       102.7  &      4.2   $\pm$     0.1  &     10.4   $\pm$     0.3  &     2.70   $\pm$    0.05  &      30.0   $\pm$     1.5  &   1 \\     
4$_{0,4}$--3$_{1,3}$     &     1279.28  &       164.2  &      4.8   $\pm$     0.3  &      8.1   $\pm$     1.0  &      1.6   $\pm$     0.4  &      13.5   $\pm$     1.2  &   1 \\     
4$_{2,3}$--4$_{1,4}$     &     1025.65  &       213.4  &      3.6   $\pm$     0.1  &     10.2   $\pm$     0.3  &     0.90   $\pm$    0.05  &       9.7   $\pm$     0.2  &   1 \\     
4$_{2,3}$--3$_{1,2}$     &     1597.81  &       213.4  &      3.9   $\pm$     0.4  &     10.5   $\pm$     1.2  &      2.6   $\pm$     0.8  &      29.4   $\pm$     2.6  &   1 \\     
4$_{4,1}$--4$_{3,2}$     &      646.87  &       281.1  &      2.8   $\pm$     0.1  &     13.8   $\pm$     0.3  &     0.36   $\pm$    0.02  &       5.3   $\pm$     0.1  &   1 \\     
6$_{0,6}$--5$_{1,5}$     &     1845.21  &       327.8  &      2.8   $\pm$     0.4  &      7.5   $\pm$     1.0  &      2.4   $\pm$     0.8  &      18.9   $\pm$     2.2  &   3 \\     
\enddata
\tablecomments{Errors for \vlsr, \dv, and \tint\ reported in the table are those computed by the Gaussian fitter within CLASS. We assume a minimum uncertainty for \vlsr\ and \dv\ of 0.1~km/s and a minimum uncertainty for \tint\ of 0.1~K km/s. The \tpeak\ error is the RMS in the local baseline, which we also measured in CLASS. Uncertainties, including additional sources of error for these measurements, are discussed in more detail in the Appendix.}
\tablenotetext{a}{Entries of 1 or 2 correspond to categories 1 (not blended) or 2 (slightly blended) as defined in Sec.~\ref{s-mli} for a particular transition, respectively. A value of 3 is given to the 6$_{0,6}$--5$_{1,5}$ transition because it is ``self-blended" with the 6$_{1,6}$--5$_{0,5}$ transition from the same \hts\ isotopologue.}
\end{deluxetable}

\clearpage

\begin{deluxetable}{lccccccc}
\tablecaption{Measured line parameters for the plateau \label{t-pl}}
\tablecolumns{8}
\tablewidth{0pt}
\tablehead{\colhead{Transition} & \colhead{$\nu$} & \colhead{E$_{up}$} & \colhead{v$_{LSR}$} & \colhead{$\Delta$v}  &
\colhead{T$_{peak}$} & \colhead{\tint} & \colhead{Notes\tablenotemark{a}} \\
\colhead{(J$_{K^{-},K^{+}}$)} & \colhead{(GHz)} & \colhead{(K)} & \colhead{(km s$^{-1}$)}   & \colhead{(km s$^{-1}$)} & \colhead{(K)} & \colhead{(K km s$^{-1}$)}  & \colhead{}}
\startdata
\cutinhead{\htsi}
2$_{0,2}$--1$_{1,1}$     &      687.30  &        54.7  &      9.7   $\pm$     0.1  &     34.5   $\pm$     0.4  &     5.46   $\pm$    0.03  &     200.4   $\pm$     1.6  &   2 \\     
2$_{1,2}$--1$_{0,1}$     &      736.03  &        55.1  &      9.8   $\pm$     0.1  &     30.6   $\pm$     0.1  &    10.62   $\pm$    0.04  &     345.7   $\pm$     0.3  &   2 \\     
2$_{2,1}$--2$_{1,2}$     &      505.57  &        79.4  &     10.7   $\pm$     0.6  &     31.6   $\pm$     0.6  &     4.57   $\pm$    0.02  &     153.7   $\pm$     1.6  &   1 \\     
2$_{2,1}$--1$_{1,0}$     &     1072.84  &        79.4  &     11.4   $\pm$     0.1  &     30.5   $\pm$     0.1  &    11.56   $\pm$    0.07  &     375.5   $\pm$     0.7  &   1 \\     
3$_{1,3}$--2$_{0,2}$     &     1002.78  &       102.9  &     10.4   $\pm$     0.1  &     29.8   $\pm$     0.1  &     7.91   $\pm$    0.1  &     250.6   $\pm$     1.4  &   2 \\     
3$_{1,2}$--2$_{2,1}$     &     1196.01  &       136.8  &     10.2   $\pm$     0.1  &     31.4   $\pm$     0.2  &      8.8   $\pm$     0.2  &     295.8   $\pm$     2.0  &   1 \\     
3$_{2,2}$--3$_{1,3}$     &      747.30  &       138.7  &      7.8   $\pm$     0.1  &     34.1   $\pm$     0.2  &     3.74   $\pm$    0.05  &     135.6   $\pm$     1.1  &   2 \\     
3$_{3,1}$--3$_{2,2}$     &      568.05  &       166.0  &      7.6   $\pm$     0.1  &     32.8   $\pm$     0.1  &     2.02   $\pm$    0.02  &      70.5   $\pm$     0.2  &   1 \\     
3$_{3,1}$--2$_{2,0}$     &     1707.97  &       166.0  &     10.0   $\pm$     0.1\tablenotemark{b}  &     36.5   $\pm$     1.2  &      4.8   $\pm$     0.6  &     186.8   $\pm$     5.0  &   1 \\     
3$_{3,0}$--2$_{2,1}$     &     1865.62  &       169.0  &     10.0   $\pm$     0.1\tablenotemark{b}  &     27.0   $\pm$     1.5  &      3.4   $\pm$     0.7  &      98.4   $\pm$     6.1  &   1 \\     
4$_{1,3}$--4$_{0,4}$     &     1018.35  &       213.2  &      7.4   $\pm$     0.2  &     31.6   $\pm$     0.6  &     3.38   $\pm$    0.09  &     113.7   $\pm$     2.3  &   1 \\     
4$_{2,3}$--3$_{3,0}$     &      930.15  &       213.6  &      7.7   $\pm$     0.1  &     31.8   $\pm$     0.7  &     0.98   $\pm$    0.06  &      33.0   $\pm$     0.7  &   2 \\     
4$_{2,3}$--4$_{1,4}$     &     1026.51  &       213.6  &      5.9   $\pm$     0.4\tablenotemark{c}  &     23.8   $\pm$     0.4  &      6.8   $\pm$     0.1  &     171.2   $\pm$     2.0  &   2 \\     
4$_{2,3}$--3$_{1,2}$     &     1599.75  &       213.6  &     11.6   $\pm$     0.7  &     34.7   $\pm$     1.3  &      3.9   $\pm$     0.6  &     143.5   $\pm$     6.3  &   1 \\     
4$_{2,2}$--4$_{1,3}$     &      665.39  &       245.1  &      9.3   $\pm$     0.5  &     32.1   $\pm$     0.5  &     2.18   $\pm$    0.05  &      74.7   $\pm$     1.4  &   2 \\     
4$_{2,2}$--3$_{3,1}$     &     1648.71  &       245.2  &      2.2   $\pm$     1.3\tablenotemark{c}  &     33.6   $\pm$     2.3  &      3.2   $\pm$     0.9  &     115.6   $\pm$    12.5  &   2 \\     
4$_{3,2}$--4$_{2,3}$     &      765.94  &       250.3  &     10.9   $\pm$     0.2  &     38.8   $\pm$     0.6  &     5.27   $\pm$    0.03  &     217.4   $\pm$     2.5  &   2 \\     
4$_{4,1}$--4$_{3,2}$     &      650.37  &       281.5  &      7.9   $\pm$     0.1  &     28.1   $\pm$     0.1  &     2.91   $\pm$    0.03  &      87.0   $\pm$     0.3  &   1 \\     
5$_{1,4}$--4$_{2,3}$     &     1852.69  &       302.5  &     11.8   $\pm$     1.3  &     32.1   $\pm$     2.2  &      2.0   $\pm$     0.7  &      68.9   $\pm$     6.3  &   1 \\     
6$_{0,6}$--5$_{1,5}$     &     1846.74  &       328.1  &     10.2   $\pm$     1.3  &     31.5   $\pm$     1.6  &      3.0   $\pm$     0.7  &      99.6   $\pm$     8.9  &   3 \\     
5$_{3,3}$--5$_{2,4}$     &     1023.12  &       351.7  &      7.6   $\pm$     0.1  &     30.8   $\pm$     0.5  &      1.9   $\pm$     0.1  &      61.1   $\pm$     0.7  &   2 \\     
5$_{3,2}$--5$_{2,3}$     &      611.44  &       379.5  &      8.3   $\pm$     0.4  &     33.5   $\pm$     1.0  &     1.41   $\pm$    0.02  &      50.2   $\pm$     1.6  &   1 \\     
5$_{4,2}$--5$_{3,3}$     &      800.85  &       390.1  &      9.5   $\pm$     0.6  &     35.7   $\pm$     0.6  &     0.85   $\pm$    0.09  &      32.2   $\pm$     1.3  &   2 \\     
6$_{5,2}$--6$_{4,3}$     &      854.97  &       558.1  &      6.9   $\pm$     0.4  &     31.1   $\pm$     0.9  &     0.90   $\pm$    0.05  &      29.9   $\pm$     0.2  &   1 \\     
6$_{5,1}$--6$_{4,2}$     &      493.36  &       563.9  &      2.4   $\pm$     0.1\tablenotemark{c}  &     18.5   $\pm$     0.3  &     0.45   $\pm$    0.01  &       8.8   $\pm$     0.2  &   1 \\     
7$_{4,3}$--7$_{3,4}$     &      880.06  &       701.1  &      2.5   $\pm$     0.2\tablenotemark{c}   &     20.0   $\pm$     0.9  &     1.52   $\pm$    0.05  &      32.5   $\pm$     1.2  &   1 \\     
\cutinhead{\htsiii}
2$_{1,2}$--1$_{0,1}$     &      734.27  &        55.0  &      7.9   $\pm$     0.1  &     33.8   $\pm$     0.6  &     2.92   $\pm$    0.04  &     105.1   $\pm$     0.9  &   2 \\     
2$_{2,1}$--1$_{1,0}$     &     1069.37  &        79.2  &      8.8   $\pm$     0.1  &     29.5   $\pm$     0.2  &     2.42   $\pm$    0.07  &      75.9   $\pm$     0.9  &   2 \\     
3$_{0,3}$--2$_{1,2}$     &      991.73  &       102.7  &      7.7   $\pm$     0.3  &     28.7   $\pm$     0.3  &     2.63   $\pm$    0.08  &      80.4   $\pm$     2.4  &   2 \\     
3$_{1,3}$--2$_{0,2}$     &     1000.91  &       102.7  &      4.6   $\pm$     0.1\tablenotemark{c}  &     29.9   $\pm$     0.4  &     1.98   $\pm$    0.08  &      62.9   $\pm$     0.6  &   2 \\     
3$_{1,2}$--2$_{2,1}$     &     1197.18  &       136.7  &      8.5   $\pm$     0.5  &     31.8   $\pm$     1.3  &      1.7   $\pm$     0.1  &      56.6   $\pm$     3.5  &   2 \\     
3$_{2,2}$--3$_{1,3}$     &      745.52  &       138.5  &      9.0   $\pm$     0.1\tablenotemark{b}  &     25.1   $\pm$     1.2  &     0.25   $\pm$    0.04  &       6.6   $\pm$     0.2  &   1 \\     
3$_{3,1}$--3$_{2,2}$     &      563.68  &       165.5  &      2.9   $\pm$     0.3\tablenotemark{c}  &     18.6   $\pm$     1.0  &     0.32   $\pm$    0.02  &       6.3   $\pm$     0.3  &   1 \\     
4$_{2,3}$--4$_{1,4}$     &     1024.85  &       213.2  &      3.2   $\pm$     0.2\tablenotemark{c}  &     22.6   $\pm$     0.7  &      1.2   $\pm$     0.1  &      29.0   $\pm$     0.9  &   2 \\     
\cutinhead{\htsii}
2$_{1,2}$--1$_{0,1}$     &      735.13  &        55.1  &      5.7   $\pm$     0.1\tablenotemark{c}  &     28.5   $\pm$     0.2  &     1.32   $\pm$    0.02  &      40.1   $\pm$     0.3  &   2 \\     
3$_{0,3}$--2$_{1,2}$     &      992.40  &       102.7  &      8.2   $\pm$     0.6  &     35.6   $\pm$     1.6  &     0.89   $\pm$    0.05  &      33.8   $\pm$     1.3  &   1 \\
\enddata
\tablecomments{Errors for \vlsr, \dv, and \tint\ reported in the table are those computed by the Gaussian fitter within CLASS. We assume a minimum uncertainty for \vlsr\ and \dv\ of 0.1~km/s and a minimum uncertainty for \tint\ of 0.1~K km/s. The \tpeak\ error is the RMS in the local baseline, which we also measured in CLASS. Uncertainties, including additional sources of error for these measurements, are discussed in more detail in the Appendix.}
\tablenotetext{a}{Entries of 1 or 2 correspond to categories 1 (not blended) or 2 (slightly blended) as defined in Sec.~\ref{s-mli} for a particular transition, respectively. A value of 3 is given to the 6$_{0,6}$--5$_{1,5}$ transition because it is ``self-blended" with the 6$_{1,6}$--5$_{0,5}$ transition from the same \hts\ isotopologue.}
\tablenotetext{b}{The \vlsr\ was held fixed during the Gaussian fitting procedure.}
\tablenotetext{c}{Indicates a \vlsr\ that is lower than typically measured toward the plateau.}
\end{deluxetable}

\clearpage

\begin{deluxetable}{lccccccc}
\tablecaption{Measured line parameters for the extended/compact ridge \label{t-er}}
\tablecolumns{8}
\tablewidth{0pt}
\tablehead{\colhead{Transition} & \colhead{$\nu$} & \colhead{E$_{up}$} & \colhead{v$_{LSR}$} & \colhead{$\Delta$v}  &
\colhead{T$_{peak}$} & \colhead{\tint} & \colhead{Notes\tablenotemark{a}} \\
\colhead{(J$_{K^{-},K^{+}}$)} & \colhead{(GHz)} & \colhead{(K)} & \colhead{(km s$^{-1}$)}   & \colhead{(km s$^{-1}$)} & \colhead{(K)} & \colhead{(K km s$^{-1}$)}  & \colhead{}}
\startdata
\cutinhead{\htsi}
2$_{0,2}$--1$_{1,1}$     &      687.30  &        54.7  &      8.5   $\pm$     0.1  &      3.2   $\pm$     0.2  &     2.25   $\pm$    0.03  &       7.6   $\pm$     0.9  &   2 \\     
2$_{1,2}$--1$_{0,1}$     &      736.03  &        55.1  &      8.7   $\pm$     0.1  &      3.3   $\pm$     0.1  &     3.35   $\pm$    0.04  &      11.7   $\pm$     0.1  &   2 \\     
2$_{2,1}$--2$_{1,2}$     &      505.57  &        79.4  &      8.6   $\pm$     0.6  &      3.1   $\pm$     0.6  &     1.95   $\pm$    0.02  &       6.4   $\pm$     1.6  &   1 \\     
2$_{2,1}$--1$_{1,0}$     &     1072.84  &        79.4  &      8.5   $\pm$     0.1  &      3.5   $\pm$     0.1  &     1.70   $\pm$    0.07  &       6.4   $\pm$     0.3  &   1 \\     
3$_{1,3}$--2$_{0,2}$\tablenotemark{d}     &     1002.78  &       102.9  &      8.8   $\pm$     0.1  &     10.9   $\pm$     0.5  &     1.56   $\pm$    0.1  &      18.1   $\pm$     1.6  &   2 \\ 
\cutinhead{\htsiii}
3$_{1,3}$--2$_{0,2}$     &     1000.91  &       102.7  &      4.8   $\pm$     0.1  &      3.4   $\pm$     0.1  &     1.97   $\pm$    0.08  &       7.2   $\pm$     0.1  &   2 \\     
\enddata
\tablecomments{Errors for \vlsr, \dv, and \tint\ reported in the table are those computed by the Gaussian fitter within CLASS. We assume a minimum uncertainty for \vlsr\ and \dv\ of 0.1~km/s and a minimum uncertainty for \tint\ of 0.1~K km/s. The \tpeak\ error is the RMS in the local baseline, which we also measured in CLASS. Uncertainties, including additional sources of error for these measurements, are discussed in more detail in the Appendix.}
\tablenotetext{a}{Entries of 1 or 2 correspond to categories 1 (not blended) or 2 (slightly blended) as defined in Sec.~\ref{s-mli} for a particular transition, respectively.}
\tablenotetext{d}{The extended ridge component for this transition is not fit well because of its relative weakness and line blending.}
\end{deluxetable}

\clearpage

\begin{deluxetable}{lccccc}
\tablecaption{Computed values for $\tau_{iso}$ and \nup(\htsi) \label{t-nup_thick}}
\tablecolumns{6}
\tablewidth{0pt}
\tablehead{\colhead{Transition} & \colhead{\tauiso} & \colhead{\nup\ (Solar)} & \colhead{\nup\ (Orion KL)} & \colhead{iso\tablenotemark{a}} \\
\colhead{(J$_{K^{-},K^{+}}$)} & \colhead{} & \colhead{($\times10^{16}$  cm$^{-2}$)}  & \colhead{($\times10^{16}$  cm$^{-2}$)} & \colhead{}}
\startdata
2$_{0,2}$--1$_{1,1}$     &   0.25     $\pm$  0.04    &    4.3     $\pm$    1.4     &    2.6   $\pm$    0.8    &     H$_{2}$$^{33}$S   \\
2$_{1,2}$--1$_{0,1}$     &   1.2       $\pm$  0.3      &    9.0    $\pm$    3.5     &    5.4    $\pm$    2.1     &     H$_{2}$$^{33}$S   \\
2$_{2,1}$--1$_{1,0}$     &   1.0       $\pm$  0.3      &    5.8    $\pm$    2.5     &    3.5    $\pm$    1.5     &     H$_{2}$$^{33}$S   \\
3$_{1,3}$--2$_{0,2}$     &   1.6       $\pm$  0.6      &    2.5    $\pm$    1.2     &    2.3    $\pm$    1.1     &     H$_{2}$$^{34}$S   \\
3$_{1,2}$--2$_{2,1}$     &   0.9       $\pm$  0.2      &    3.2    $\pm$    1.4     &    2.9    $\pm$    1.3     &     H$_{2}$$^{34}$S   \\
3$_{2,2}$--3$_{1,3}$     &   0.45     $\pm$  0.08    &    2.1    $\pm$    0.7      &    1.9    $\pm$    0.6    &     H$_{2}$$^{34}$S   \\
3$_{3,1}$--3$_{2,2}$     &   0.24     $\pm$  0.04    &    0.7    $\pm$    0.2      &    0.6    $\pm$    0.2    &     H$_{2}$$^{34}$S   \\
3$_{3,1}$--2$_{2,0}$     &   0.9       $\pm$  0.4      &    1.0    $\pm$    0.8     &    0.9    $\pm$    0.7     &     H$_{2}$$^{34}$S   \\
4$_{2,3}$--4$_{1,4}$     &   0.5       $\pm$  0.1      &    1.7    $\pm$    0.7     &    1.5    $\pm$    0.6     &     H$_{2}$$^{34}$S   \\
4$_{2,3}$--4$_{1,4}$     &   0.23     $\pm$  0.04    &    4.3     $\pm$    1.6      &    2.6   $\pm$    1.0   &     H$_{2}$$^{33}$S   \\
4$_{2,3}$--3$_{1,2}$     &   1.8      $\pm$  1.1      &    3.6     $\pm$    3.1      &    3.3   $\pm$    2.8   &     H$_{2}$$^{34}$S   \\
4$_{2,3}$--3$_{1,2}$     &   0.5      $\pm$  0.2      &    5.9     $\pm$    4.3      &    3.6   $\pm$    2.6   &     H$_{2}$$^{33}$S   \\
4$_{2,2}$--4$_{1,3}$     &   0.38    $\pm$  0.07    &    1.2      $\pm$    0.4      &    1.1   $\pm$    0.3   &     H$_{2}$$^{34}$S   \\
4$_{2,2}$--3$_{3,1}$     &   0.3      $\pm$  0.2      &    1.7     $\pm$    1.3     &    1.5    $\pm$    1.1    &     H$_{2}$$^{34}$S   \\
4$_{4,1}$--4$_{3,2}$     &   0.7      $\pm$  0.1      &    2.3     $\pm$    0.8     &    2.1    $\pm$    0.7    &     H$_{2}$$^{34}$S   \\
4$_{4,1}$--4$_{3,2}$     &   0.17    $\pm$  0.03    &    3.0     $\pm$    0.9      &    1.8    $\pm$    0.6    &     H$_{2}$$^{33}$S   \\
5$_{1,4}$--4$_{2,3}$     &   1.0      $\pm$  0.5      &    2.0     $\pm$    1.8     &    1.8    $\pm$    1.6    &     H$_{2}$$^{34}$S   \\
5$_{2,4}$--4$_{1,3}$     &   0.6      $\pm$  0.3      &    1.3    $\pm$    1.1     &    1.2     $\pm$    1.0     &     H$_{2}$$^{34}$S   \\
5$_{3,3}$--5$_{2,4}$     &   0.22    $\pm$  0.04    &    0.6    $\pm$    0.2     &    0.6      $\pm$    0.2     &     H$_{2}$$^{34}$S   \\
5$_{3,2}$--5$_{2,3}$     &   0.48    $\pm$  0.09    &    1.3    $\pm$    0.4     &    1.2      $\pm$    0.4     &     H$_{2}$$^{34}$S   \\
5$_{4,2}$--5$_{3,3}$     &   0.16    $\pm$  0.03    &    0.4    $\pm$    0.1     &    0.3      $\pm$    0.1     &     H$_{2}$$^{34}$S   \\
6$_{3,3}$--6$_{2,4}$     &   0.13    $\pm$  0.03    &    0.24   $\pm$    0.08   &    0.22    $\pm$    0.08   &     H$_{2}$$^{34}$S   \\
6$_{5,2}$--6$_{4,3}$     &   0.14    $\pm$  0.03    &    0.3     $\pm$    0.1     &    0.3     $\pm$    0.1    &     H$_{2}$$^{34}$S   \\
7$_{4,3}$--7$_{3,4}$     &   0.27    $\pm$  0.05    &    0.17   $\pm$    0.06   &    0.15    $\pm$    0.05   &     H$_{2}$$^{34}$S   \\
\enddata
\tablecomments{Uncertainty calculations for \tauiso\ and \nup\ are described in the Appendix.}
\tablenotetext{a}{This column indicates the isotopologue used to compute $\tau_{iso}$.}
\end{deluxetable}

\clearpage

\begin{deluxetable}{ccc}
\tablecaption{Ortho/Para Ratio Estimates \label{t-op}}
\tablecolumns{3}
\tablewidth{0pt}
\tablehead{\colhead{Ortho State} & \colhead{Para State} & \colhead{Ortho/Para Ratio}}
\startdata
2$_{1,2}$  &  2$_{0,2}$  &  2.1 $\pm$ 1.1 \\ 
3$_{1,2}$  &  3$_{2,2}$  &  1.5 $\pm$ 0.8 \\
5$_{1,4}$  &  5$_{2,4}$  &  1.6 $\pm$ 1.9 \\ 
\enddata
\tablecomments{The first two columns indicate the ortho and para state from which an \nup\ estimate was taken in order to compute the ortho/para ratio in the third column.}
\end{deluxetable}

\clearpage

\begin{deluxetable}{lcccc}
\tablecaption{Transitions Used for \ntot(HDS) Upper Limit \label{t-hds}}
\tablecolumns{6}
\tablewidth{0pt}
\tablehead{\colhead{Transition} & \colhead{$\nu_{rest}$} & \colhead{E$_{up}$} & \colhead{$\mu^{2}$S} & \colhead{Local RMS\tablenotemark{a}} \\
\colhead{(J$_{K^{-},K^{+}}$)} & \colhead{(GHz)} & \colhead{(K)} & \colhead{(D$^{2}$)} & \colhead{(mK)}}
\startdata
2$_{2,0}$--1$_{1,1}$     &      1035.61  &      68.4  &     0.75  &     84  \\ 
3$_{1,3}$--2$_{0,2}$     &       754.96  &       70.9  &     1.27  &     45  \\ 
3$_{0,3}$--2$_{1,2}$     &       586.90  &       67.8  &     1.01  &     28  \\ 
3$_{2,2}$--2$_{1,1}$     &     1166.00  &     103.0  &     0.99  &     150 \\ 
3$_{2,2}$--2$_{2,1}$     &       732.73  &     103.0  &     0.60  &      22   \\ 
4$_{1,4}$--3$_{0,3}$     &       920.50  &     112.0  &     1.77  &      47 \\ 
\enddata
\tablenotetext{a}{The RMS was computed using CLASS in the vicinity of the rest frequency.}
\end{deluxetable}

\clearpage

\bibliographystyle{apj}
\bibliography{ms.bbl}

\end{document}